\documentclass[journal,12pt,onecolumn]{IEEEtran}
\usepackage{setspace}
\ifCLASSOPTIONonecolumn \doublespace \fi

\usepackage{graphicx}
\graphicspath{{Figures/}} \DeclareGraphicsExtensions{.eps,.ps}

\usepackage{amsfonts}
\usepackage{amsmath}
\usepackage{cite}
\allowdisplaybreaks[1]

\ifCLASSINFOpdf
\else
\fi

\begin{document}
%
\title{A Novel Hybrid Beamforming Algorithm with Unified Analog Beamforming by Subspace Construction Based on Partial CSI for Massive MIMO-OFDM Systems}
%
%
%

\author{Dengkui~Zhu,
        Boyu~Li,~\IEEEmembership{Member,~IEEE,}
        and~Ping~Liang~\IEEEmembership{}
\thanks{D. Zhu and B. Li are with RF DSP Inc., Irvine, CA 92606, USA (e-mail:
byli@rfdsp.com; dkzhu@rfdsp.com).}
\thanks{
P. Liang is with the University of California {-} Riverside, Riverside, CA
92521, USA, and also with RF DSP Inc., Irvine, CA 92606, USA (e-mail:
liang@ee.ucr.edu).}
}

\maketitle

\begin{abstract}
Hybrid beamforming (HB) has been widely studied for reducing the number of costly radio frequency (RF) chains in massive multiple-input multiple-output (MIMO) systems. However, previous works on HB are limited to a single user equipment (UE) or a single group of UEs, employing the frequency-flat first-level analog beamforming (AB) that cannot be applied to multiple groups of UEs served in different frequency resources in an orthogonal frequency-division multiplexing (OFDM) system. In this paper, a novel HB algorithm with unified AB based on the spatial covariance matrix (SCM) knowledge of all UEs is proposed for a massive MIMO-OFDM system in order to support multiple groups of UEs. The proposed HB method with a much smaller number of RF chains can achieve more than $95\%$ performance of full digital beamforming. In addition, a novel practical subspace construction (SC) algorithm based on partial channel state information is proposed to estimate the required SCM. The proposed SC method can offer more than $97\%$ performance of the perfect SCM case. With the proposed methods, significant cost and power savings can be achieved without large loss in performance. Furthermore, the proposed methods can be applied to massive MIMO-OFDM systems in both time-division duplex and frequency-division duplex.
\end{abstract}


%

\section{Introduction} \label{sec:introduction}
\ifCLASSOPTIONonecolumn In \else \IEEEPARstart{I}n \fi the past few years, the massive Multiple-Input Multiple-Output (MIMO) technique \cite{Marzetta_massive_MIMO_original, Rusek_massive_MIMO_overview, Hoydis_massive_MIMO, Larsson_massive_MIMO_overview} 
has been considered as one of the most promising candidates for the Fifth Generation (5G) standard of mobile communication, and, it is being standardized by the Third Generation Partnership Project (3GPP) \cite{3GPP_TR_36.897}. 
In massive MIMO systems, each Base-Station (BS) is equipped with tens to hundreds of antennas, where each BS antenna is connected to its own  Radio Frequency (RF) chain, to serve tens of User Equipments (UEs) simultaneously in the same time-frequency resource
. In this way, tremendous advantages could be achieved. 
One of the most important advantages is that such systems can simultaneously boost capacity and energy efficiency greatly due to its capability to achieve a very large array gain and aggressive spatial multiplexing at the same time. In general, Time-Division Duplex (TDD) is considered for such systems, exploiting the channel reciprocity between the uplink and downlink \cite{Marzetta_massive_MIMO_original, Rusek_massive_MIMO_overview, Hoydis_massive_MIMO, Larsson_massive_MIMO_overview}
, although Frequency-Division Duplex (FDD) has also been considered \cite{Adhikary_JSDM,Jiang_FDD_Massive_MIMO,PCT_US2014_071752}.

Despite the theoretical advantages, implementing a large number of RF chains in a massive
MIMO system can be problematic, since it increases the system cost and power consumption, and lowers power efficiency. To address these issues, Hybrid Beamforming (HB) \cite{Wu_HB,Ayach_mmWave,Alkhateeb_HB_CE,Geng_MUHB,Liang_HB,
Bogale_MUHB,Chiu_HB,Xu_HB,Ying_HB,Stirling_MUHB,Alkhateeb_HP,Sohrabi_HB,Sohrabi_HB_OFDM} 
has been proposed
. HB is realized by two-level beamforming. Specifically, the high-dimensional first-level 
RF Analog Beamforming (AB), which could be realized by a low-cost Phase Shift Network (PSN), is applied to decrease the number of RF chains, before the reduced-dimensional second-level digital baseband beamforming is employed. 

There has been increasing interest in HB for different application scenarios. In this paper, we focus on the important application of HB in a massive Multi-User MIMO (MU-MIMO) Orthogonal Frequency-Division Multiplexing (OFDM) system that serves multiple groups of UEs with each group being served by a different frequency resource. We further focus on sub $6\mathrm{GHz}$ frequency bands used by cellular network today and will most likely still used in the future. Previous works on HB are not well suited for such applications because the limitations below.
\begin{enumerate}
\item{In sub $6\mathrm{GHz}$ bands, HB methods were proposed for Single-User MIMO (SU-MIMO) systems in \cite{Xu_HB,Sohrabi_HB,Sohrabi_HB_OFDM}
, and for MU-MIMO systems in \cite{Geng_MUHB,Liang_HB,Bogale_MUHB,Ying_HB,Sohrabi_HB}
. However, their frequency-flat first-level AB methods were designed for a single UE or a single group of UEs, where all UEs in the group are served on the same frequency resource. They 
cannot be applied in the case of multiple groups of UEs with each group being served by a different frequency resource. A massive MIMO HB method supporting multiple groups of UEs offers advantages over methods that only support a single UE or a single group of UEs. Prior HB methods that allocate the whole band to a single UE or a single group of UEs suffer from low flexibility in UE scheduling, which may cause high latency of data in UEs due to the
relatively small number of scheduled UEs per scheduling period, and low resource use efficiency because some UEs do not need to be served by the whole band.}

\item{Even if the whole frequency band is allocated for a single UE or a single group of UEs
, challenges still exist to apply HB algorithms in \cite{Geng_MUHB,Liang_HB,Xu_HB,Bogale_MUHB,Ying_HB,Sohrabi_HB,Sohrabi_HB_OFDM}   
 to the considered massive MIMO-OFDM systems
 . Specifically, full Channel State Information (CSI) is assumed 
  at either each BS or each UE, which is challenging to be acquired with HB because the number of RF chains is much smaller than the number of antennas. In the ideal case where each RF chain is connected to all antennas as in\cite{Ayach_mmWave,Alkhateeb_HB_CE,Liang_HB,
Bogale_MUHB,Ying_HB,Alkhateeb_HP,Sohrabi_HB,Sohrabi_HB_OFDM}
, although it is possible to estimate full CSI in succession, it would significantly increase the channel estimation overhead and system complexity, which have already caused difficulties in realizing aggressive spatial multiplexing. In addition, in the more practical case where each RF chain is only connected to a subset of antennas as in \cite{Wu_HB,Geng_MUHB,Chiu_HB,Xu_HB,Stirling_MUHB,Sohrabi_HB_OFDM}
, it is not even possible to acquire full CSI. 
}

\item{HB algorithms proposed in \cite{Ayach_mmWave,Alkhateeb_HB_CE} 
 for SU-MIMO systems, and in \cite{Wu_HB,Chiu_HB,Stirling_MUHB,Alkhateeb_HP} for MU-MIMO systems, are specifically designed for millimeter Wave (mmWave) systems where the number of multipath components is very limited and Line-of-Sight (LoS) channels are generally assumed. They cannot be directly applied to sub $6\mathrm{GHz}$ frequency bands where many multipath components exist and Non-LoS (NLoS) channels are very common.}
\end{enumerate} 

In this paper, we propose a novel HB method with unified AB by Subspace Construction (SC) based on partial CSI that overcomes the limitations identified above
. The contributions are summarized below.
\begin{enumerate}
\item{A novel HB algorithm with unified AB 
is particularly designed for massive MIMO-OFDM systems in sub $6\mathrm{GHz}$ frequency bands 
to support MU-MIMO beamforming for multiple groups of UEs, which cannot be achieved by previous works on HB as discussed above
. Instead of full CSI assumed by prior HB methods, the proposed HB method is based on the second-order Spatial Covariance Matrix (SCM), which is a much more practical assumption than full CSI because it changes much slower. The sufficient number of RF chains is derived for the proposed HB to offer the performance close to complete digital beamforming without a reduced number of RF chains. Simulation results show that our proposed HB method with a reduced number of RF chains can achieve no less than $95{\%}$ performance of complete digital beamforming without the reduced number of RF chains.}

\item{Although the change of the SCM is much slower than the instantaneous CSI, it is still based on full CSI. As identified earlier, in the case of HB, since the number of BS antennas is much larger than the number of RF chains, it is still a challenge to efficiently acquire the SCM. To tackle this issue, in this paper, we develop a novel SC algorithm based on only partial CSI to estimate the required SCM 
for the proposed HB in practice. Simulation results show that our proposed SC method can offer more than $97\%$ performance of the perfect SCM case. Our proposed SC method may be used in other applications such as \cite{Adhikary_JSDM,Jiang_FDD_Massive_MIMO}. With the proposed HB and SC methods, significant cost and power savings can be achieved without large performance loss in practice.}

\item{The proposed HB and SC algorithms can be applied in many scenarios. Particularly, their applications in both TDD and FDD systems are discussed in details. In addition, although they are designed for multiple groups of UEs in OFDM systems in this paper, the proposed algorithms can be easily applied to a single UE or a single group of UEs for both OFDM and single-carrier systems. Moreover, since they can be applied to both LoS and NLoS channels, with simple generalization to the case of multiple-antenna UEs, it promises to be applicable to mmWave systems, which is better dealt in a separate paper, hence is not discussed in detail in this paper.}
\end{enumerate}

The remainder of this paper is organized as follows. The considered system and channel models are first presented in Section \ref{sec:models}. In Section \ref{sec:hybrid}, the proposed HB with unified AB is provided and the sufficient number of RF chains is derived for the proposed HB to achieve the performance close to complete digital beamforming without a reduced number of RF chains. Then, the proposed SC method based on only partial instantaneous CSI is proposed in Section \ref{sec:subspace}. In Section \ref{sec:fdd}, the applications of the proposed HB and SC algorithms in FDD systems are discussed. After that, simulation results are provided in Section{\ref{sec:results}. Finally, conclusions are drawn in Section \ref{sec:conclusions}.

\section{System and Channel Models} \label{sec:models}
\subsection{System Model} \label{subsec:system}
Consider an OFDM-based massive MIMO wireless communication system in sub $6\mathrm{GHz}$ where the BS is equipped with $R$ RF chains connecting to $M$ antennas where $M$ is large, e.g., $256$, and $R\ll M$. For OFDM, $L$-point Fast Fourier Transform (FFT) is employed, and $L_{\mathrm{used}} <= L$ subcarriers are used for data transmission. For each time-frequency resource, e.g., a Resource Block (RB) in the 3GPP Long Term Evolution (LTE) standard \cite{3GPP_TS_36.201}, either SU-MIMO beamforming or MU-MIMO beamforming can be applied. In this paper, for the sake of simplicity, at the $l$th subcarrier, for a time period within the channel coherence time, $K_{l}\leq R$ single-antenna UEs are assumed to be simultaneously served and the frequency-domain $M \times K_{l}$ channel matrix between BS antennas and UEs for the uplink is denoted by $\mathbf{H}_l$, where $l=1,2,\ldots,L$. Note that single-antenna UEs are assumed in this paper for the sake of clarity, which can be easily generalized to the case of multiple-antenna UEs as long as antennas at the same UE are not strongly correlated. The number of $K_l$ can be different for different subcarriers, and the same UE can be included in different served UE groups at different subcarriers. Unless otherwise specified, TDD is assumed in this paper to exploit channel reciprocity between the uplink and downlink. In our case, for TDD, the frequency-domain channel matrix for the downlink at the $l$th subcarrier is then $\mathbf{H}_l^{\mathrm{T}}$. Since the system models for the uplink and downlink are symmetric, this paper focuses only on the downlink, but the contents can be easily generalized to the uplink.  

For the downlink, at the $l$th subcarrier, the $K_l \times 1$ modulated symbol vector $\mathbf{s}_l$, where $\mathrm{E}[\mathbf{s}_l \mathbf{s}^{\mathrm{H}}_l ] = 1/K_l \mathbf{I}_{K_l}$ with $\mathbf{I}_{X}$ being the $X$-dimensional identity matrix, is first precoded by the $R \times K_l$ frequency-domain baseband digital precoding matrix $\mathbf{W}^{\mathrm{BB}}_l$ as 
\begin{align}
\mathbf{x}^{\mathrm{BB}}_l = \mathbf{W}^{\mathrm{BB}}_l \mathbf{s}_l,
\label{eq:precoding_bb}
\end{align}
where $\mathbf{x}^{\mathrm{BB}}_l$ is the $R \times 1$ precoded baseband symbol vector. Note that although only $L_{\mathrm{used}}$ subcarriers are used
, for the sake of simplicity, all subcarriers are formulated in the same way only with $0<K_l\leq R$ for used subcarriers 
while $K_l=0$ for unused subcarriers
. Next, the frequency-domain symbols are transformed into the time domain by Inverse FFT (IFFT) for each RF chain and Cyclic Prefix (CP) is added, which are the same as conventional OFDM systems. Then, the $M \times R$ time-domain unified RF analog precoding matrix $\mathbf{W}^{\mathrm{RF}}$ is applied to the time-domain symbols. 
In this paper, unless otherwise specified, it is assumed that $\mathbf{W}^{\mathrm{RF}}$ can only adjust phases, which is generally realized by a PSN. Since 
$\mathbf{W}^{\mathrm{RF}}$ has no frequency selectivity, it is equivalent to being applied in the frequency domain. As a result, at the $l$th subcarrier, the $M \times 1$ frequency-domain transmitted RF symbol vector 
can be represented as
\begin{align}
\mathbf{x}^{\mathrm{RF}}_l = \mathbf{W}^{\mathrm{RF}} \mathbf{x}^{\mathrm{BB}}_l = \mathbf{W}^{\mathrm{RF}} \mathbf{W}^{\mathrm{BB}}_l \mathbf{s}_l = \mathbf{W}_l \mathbf{s}_l,
\label{eq:precoding}
\end{align}  
where $\mathbf{W}_l =  \mathbf{W}^{\mathrm{RF}} \mathbf{W}^{\mathrm{BB}}_l$ is the $M\times K_l$ frequency-domain effective precoding matrix for 
$\mathbf{s}_l$. In this paper, the transmitting power is under the sum power constraint of $1$, so $||\mathbf{W}_l||^2_{\mathrm{F}} = 1/L_{\mathrm{used}}$ for the $L_{\mathrm{used}}$ used subcarriers 
where $||\cdot||_{\mathrm{F}}$ denotes the Frobenius norm. 
The transmitter structure is similar to \cite{Sohrabi_HB_OFDM}.


At UEs, after removing CP and FFT, the received time-domain symbol is converted back into the frequency domain. For the $l$th subcarrier, the frequency-domain input-output relation is 
\ifCLASSOPTIONonecolumn
\begin{align}
\mathbf{y}_l = \mathbf{H}_l^{\mathrm{T}} \mathbf{x}^{\mathrm{RF}}_l + \mathbf{n}_l = \mathbf{H}_l^{\mathrm{T}} \mathbf{W}^{\mathrm{RF}} \mathbf{W}^{\mathrm{BB}}_l \mathbf{s}_l + \mathbf{n}_l = \mathbf{H}_l^{\mathrm{T}} \mathbf{W}_l \mathbf{s}_l + \mathbf{n}_l,
\label{eq:io}
\end{align}  
\else
\begin{align}
\mathbf{y}_l = \mathbf{H}_l^{\mathrm{T}} \mathbf{x}^{\mathrm{RF}}_l + \mathbf{n}_l 
= \mathbf{H}_l^{\mathrm{T}} \mathbf{W}_l \mathbf{s}_l + \mathbf{n}_l,
\label{eq:io}
\end{align}  
\fi
where $\mathbf{y}_l$ and $\mathbf{n}_l$ are the $K_l\times 1$ frequency-domain received symbol vector and noise vector with each element being a zero-mean complex-valued Gaussian random variable, for the $K_l$ related UEs respectively. The input-output relation (\ref{eq:io}) can be comprehended in another way as
\begin{align}
\mathbf{y}_l = \mathbf{H}_l^{\mathrm{T}} \mathbf{W}^{\mathrm{RF}} \mathbf{x}^{\mathrm{BB}}_l + \mathbf{n}_l = \left(\mathbf{H}^{\mathrm{BB}}_l\right)^{\mathrm{T}} \mathbf{x}^{\mathrm{BB}}_l + \mathbf{n}_l,
\label{eq:io_bb}
\end{align} 
where $(\mathbf{H}_l^{\mathrm{BB}})^{\mathrm{T}} = \mathbf{H}_l^{\mathrm{T}} \mathbf{W}^{\mathrm{RF}}$ is the $K_l \times R$ effective frequency-domain downlink channel matrix for 
$\mathbf{x}^{\mathrm{BB}}_l$. Note that 
to spatially multiplex $K_l$ streams, 
$K_l\leq R$ needs to be satisfied. 

\subsection{Channel Model} \label{subsec:channel}
In this paper, the Three-Dimensional (3D) geometrically based statistical channel model for Uniform Planar Array (UPA) employed in \cite{3GPP_TR_36.873,Ayach_mmWave} is considered. Fig. \ref{fig:upa} illustrates a UPA based on spherical coordinates. The UPA consists $M=M_{\mathrm{h}} \times M_{\mathrm{v}}$ antennas where each row has $M_{\mathrm{h}}$ antennas and each column has $M_{\mathrm{v}}$ antennas. The horizontal and vertical distances between two neighboring antennas are $d_{\mathrm{h}}$ and $d_{\mathrm{v}}$ respectively. For a direction vector $\vec{v}$, the symbols $\phi$ and $\theta$ denote the azimuth and elevation angles respectively 
where $\phi, \theta \in [0,\pi]$. 

\ifCLASSOPTIONonecolumn
\begin{figure}[!t]
\centering \includegraphics[width = 0.65\linewidth]{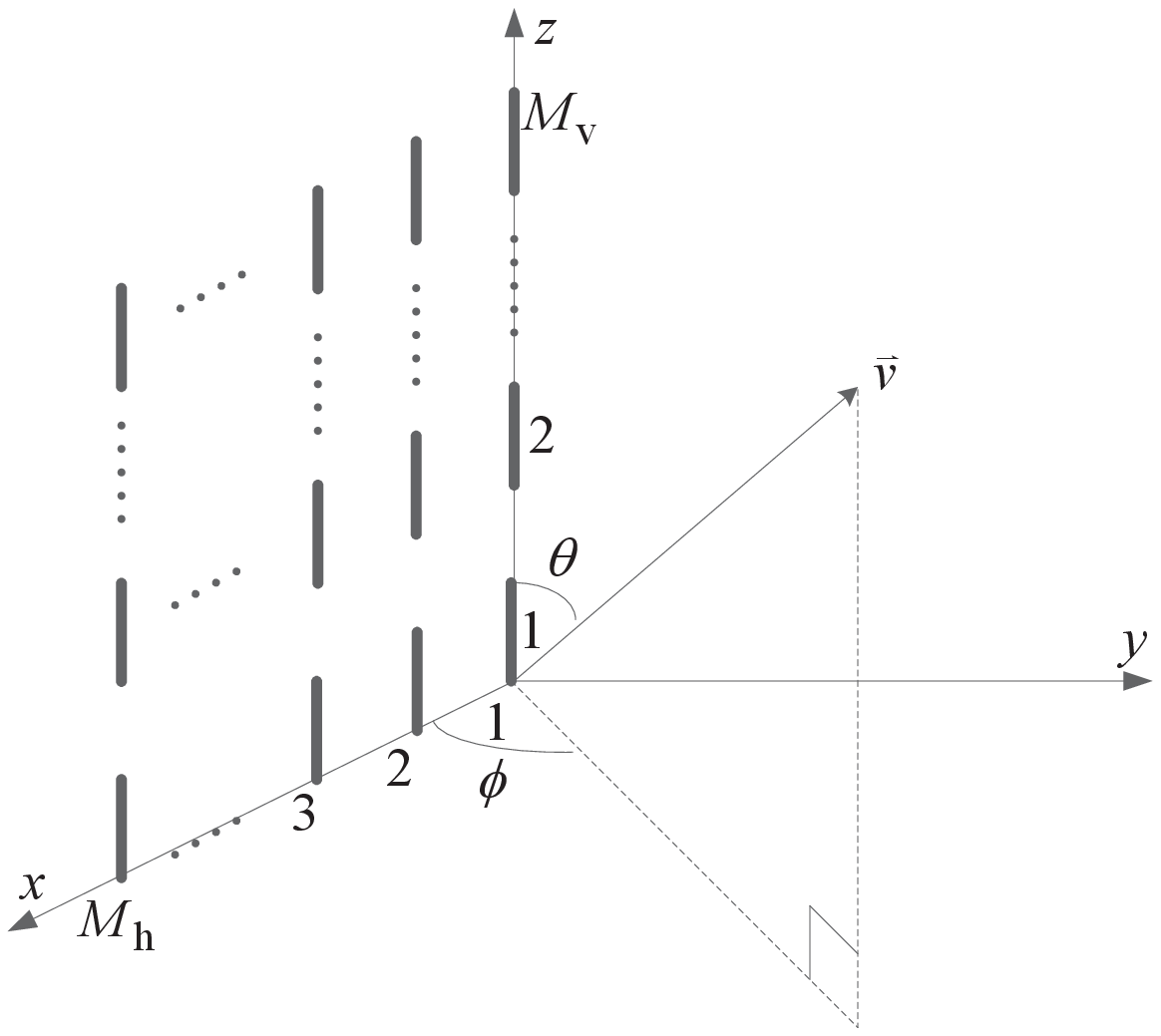}
\caption{A UPA based on spherical coordinates.}
\label{fig:upa}
\end{figure}
\else
\begin{figure}[!t]
\centering \includegraphics[width = 0.8\linewidth]{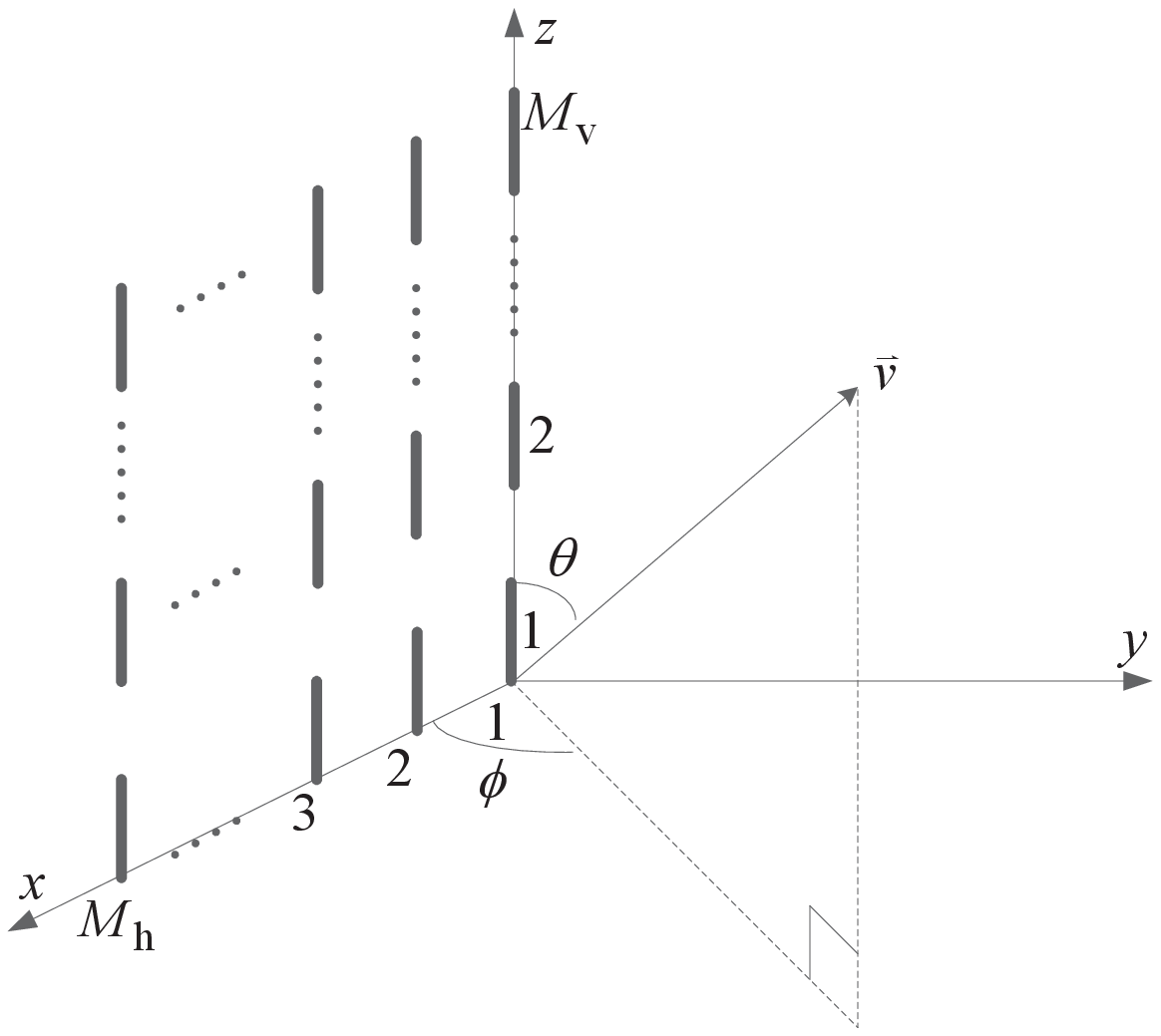}
\caption{A UPA based on spherical coordinates.}
\label{fig:upa}
\end{figure}
\fi

Let $K$ denote the number of served single-antenna UEs for the whole band within the channel coherence time, where $K_l \leq K \leq \sum_{l=1}^L K_l$. Then, for the $k$th served UE where $k=1,\ldots,K$, the discrete-time $M \times 1$ normalized channel vector can be modeled as the sum of contributions of $N_{\mathrm{cl},k}$ scattering clusters, each of which contributes $N_{\mathrm{ray},k}$ propagation rays, as
\ifCLASSOPTIONonecolumn
\begin{align}
\tilde{\mathbf{h}}_k\left(n\right) = \sqrt{\frac{M}{N_{\mathrm{cl},k}N_{\mathrm{ray},k}}}\sum_{i=1}^{N_{\mathrm{cl},k}}\sum_{j=1}^{N_{\mathrm{ray},k}}\tilde{\alpha}_{ijk}\Delta\left(\phi_{ijk},\theta_{ijk}\right)\mathbf{a}\left(\phi_{ijk},\theta_{ijk}\right)\delta\left(n-\tau_{ijk}\right),
\label{eq:channel_time}
\end{align} 
\else
\begin{align}
\tilde{\mathbf{h}}_k\left(n\right) = & \sqrt{\frac{M}{N_{\mathrm{cl},k}N_{\mathrm{ray},k}}}\sum_{i=1}^{N_{\mathrm{cl},k}}\sum_{j=1}^{N_{\mathrm{ray},k}}\tilde{\alpha}_{ijk}\Delta\left(\phi_{ijk},\theta_{ijk}\right)\nonumber \\
& \times \mathbf{a}\left(\phi_{ijk},\theta_{ijk}\right)\delta\left(n-\tau_{ijk}\right),
\label{eq:channel_time}
\end{align} 
\fi
where $\tilde{\alpha}_{ijk}$ and $\tau_{ijk}$ are the complex gain and the number of delayed samples of the $j$th ray in the $i$th scattering cluster respectively, $\phi_{ijk}$ and $\theta_{ijk}$ are the azimuth and elevation angles of arrival or departure for the uplink or downlink of the $j$th ray in the $i$th scattering cluster at the BS respectively, and $\Delta(\phi,\theta)$ and $\mathbf{a}(\phi,\theta)$ denote the antenna element gain and the normalized antenna array response vector corresponding to $\phi$ and $\theta$ respectively. In this paper, $\tilde{\alpha}_{ijk}$ and $\tau_{ijk}$ are modeled based on \cite{3GPP_TR_36.873}. For the sake of simplicity, BS antennas are modeled as ideal sector antennas \cite{Ayach_mmWave} so that  
\begin{equation}
\Delta\left(\phi,\theta\right) = \left\lbrace 
\begin{array}{cc}
1, & \phi \in \left[\phi_{\mathrm{min}},\phi_{\mathrm{max}} \right] \textrm{ and } \theta \in \left[\theta_{\mathrm{min}},\theta_{\mathrm{max}} \right], \\
0, & \textrm{otherwise},
\end{array} \right.
\label{eq:ant_gain}
\end{equation}
where $\phi_{\mathrm{min}}$, $\phi_{\mathrm{max}}$, $\theta_{\mathrm{min}}$, and $\theta_{\mathrm{max}}$ define the sector for BS antennas. As for UE antennas, for the sake of simplicity, ideal omnidirectional antennas are assumed, i.e., antenna gains are $1$ for all directions, so that they are not formulated in (\ref{eq:channel_time}). For the xz-plane UPA presented in Fig. \ref{fig:upa}, the normalized antenna array response vector can be written as 
\begin{align}
\mathbf{a}\left(\phi,\theta\right) = \mathbf{a}_{\mathrm{h}}\left(\phi,\theta\right) \otimes  \mathbf{a}_{\mathrm{v}}\left(\theta\right),
\label{eq:upa}
\end{align}
with $\otimes$ denoting the Kronecker product, where
\ifCLASSOPTIONonecolumn
\begin{align}
\mathbf{a}_{\mathrm{h}}\left(\phi,\theta\right) = \frac{1}{\sqrt{M_{\mathrm{h}}}}\left[1, \,\, e^{j2\pi \frac{d_{\mathrm{h}}}{\lambda}\cos\phi\sin\theta}, \,\, \cdots, \,\, e^{j2\pi \left(M_{\mathrm{h}}-1\right)\frac{d_{\mathrm{h}}}{\lambda}\cos\phi\sin\theta}\right]^{\mathrm{T}},
\label{eq:upa_horizontal}
\end{align}
\else
\begin{align}
\mathbf{a}_{\mathrm{h}}\left(\phi,\theta\right) = & \frac{1}{\sqrt{M_{\mathrm{h}}}}\left[1, \,\, e^{j2\pi \frac{d_{\mathrm{h}}}{\lambda}\cos\phi\sin\theta}, \,\, \right. \nonumber \\
& \left.  \cdots, \,\, e^{j2\pi \left(M_{\mathrm{h}}-1\right)\frac{d_{\mathrm{h}}}{\lambda}\cos\phi\sin\theta}\right]^{\mathrm{T}},
\label{eq:upa_horizontal}
\end{align}
\fi
and
\begin{align}
\mathbf{a}_{\mathrm{v}}\left(\theta\right) = \frac{1}{\sqrt{M_{\mathrm{v}}}}\left[1, \,\, e^{j2\pi \frac{d_{\mathrm{v}}}{\lambda}\cos\theta}, \,\, \cdots, \,\, e^{j2\pi \left(M_{\mathrm{v}}-1\right)\frac{d_{\mathrm{v}}}{\lambda}\cos\theta}\right]^{\mathrm{T}},
\label{eq:upa_vertical}
\end{align}
with $\lambda$ being the wavelength
. Since single-antenna UEs are assumed in this paper, the normalized antenna array response vector for each UE is degenerated to $1$, hence it is not included in (\ref{eq:channel_time}). Note that the xz-plane UPA considered in this subsection is just an arbitrary choice to present an example, and all results in this paper hold regardless of array placement. In addition, when $M_{\mathrm{h}}=1$ or $M_{\mathrm{v}}=1$, (\ref{eq:upa})-(\ref{eq:upa_vertical}) is degenerated to Uniform Linear Array (ULA), so all results in this paper can be applied to ULA as well. 

After FFT, at the $l$th subcarrier, the frequency-domain counterpart of $\tilde{\mathbf{h}}_{\mathrm{k}}(n)$ in (\ref{eq:channel_time}) can be written as
\begin{align}
\mathbf{h}_{kl} = \sqrt{\frac{M}{N_{\mathrm{cl},k}N_{\mathrm{ray},k}}}\sum_{i=1}^{N_{\mathrm{cl},k}}\sum_{j=1}^{N_{\mathrm{ray},k}}\alpha_{ijkl}\mathbf{a}\left(\phi_{ijk},\theta_{ijk}\right),
\label{eq:channel_freq}
\end{align} 
where $\alpha_{ijkl} = \tilde{\alpha}_{ijk}e^{-j2\pi\tau_{ijk}(l-1)/L}$. This equation 
indicates that 
$\mathbf{h}_{kl}$ at any subcarrier is located in the same spatial subspace constructed by $\mathbf{a}(\phi_{ijk},\theta_{ijk})$, which is an $M\times (N_{\mathrm{cl},k}N_{\mathrm{ray},k})$ matrix denoted by
\ifCLASSOPTIONonecolumn
\begin{align}
\mathbf{V}_k = \mathrm{span}\left\lbrace \sqrt{\left|\tilde{\alpha}_{ijk}\right|} \mathbf{a}_\mathrm{h}\left(\phi_{ijk},\theta_{ijk}\right) \right\rbrace \otimes \mathrm{span}\left\lbrace \sqrt{\left|\tilde{\alpha}_{ijk}\right|}\mathbf{a}_\mathrm{v}\left(\theta_{ijk}\right) \right\rbrace, \,\, \forall i,j.
\label{eq:subspace}
\end{align} 
\else
\begin{align}
\mathbf{V}_k = & \mathrm{span}\left\lbrace \sqrt{\left|\tilde{\alpha}_{ijk}\right|} \mathbf{a}_\mathrm{h}\left(\phi_{ijk},\theta_{ijk}\right) \right\rbrace \nonumber \\ 
& \otimes \mathrm{span}\left\lbrace \sqrt{\left|\tilde{\alpha}_{ijk}\right|}\mathbf{a}_\mathrm{v}\left(\theta_{ijk}\right) \right\rbrace, \,\, \forall i,j.
\label{eq:subspace}
\end{align} 
\fi
Note that in (\ref{eq:channel_freq}), all rays of all clusters are assumed to be located in $[\phi_{\mathrm{min}},\phi_{\mathrm{max}}]$ and $[\theta_{\mathrm{min}},\theta_{\mathrm{max}}]$, so the parameter $\Delta(\phi_{ijk},\theta_{ijk})$ in (\ref{eq:channel_time}) is omitted.
 
\section{A Novel HB Algorithm with Unified AB} \label{sec:hybrid}

\subsection{Design Problem Formulation} \label{subsec:problem}

For massive MIMO systems, 
the Zero-Forcing (ZF) precoding 
can achieve the sum rate very close to the capacity-achieving Dirty Paper Coding (DPC) under favorable propagation, hence it is virtually optimal \cite{Rusek_massive_MIMO_overview}. Since 
the linear ZF precoding requires significantly less complexity than the nonlinear DPC
, it has been considered as one of the potential practical precoding methods for massive MIMO systems \cite{Rusek_massive_MIMO_overview}
.
Hence, it 
is considered as the benchmark method in this paper, which is the same as in \cite{Liang_HB,Ying_HB,Alkhateeb_HP,Sohrabi_HB}. Specifically, the 
 ZF precoding for massive MIMO system is 
\begin{align}
\mathbf{W}^{\mathrm{ZF}}_l = \beta_l \left(\mathbf{H}_l^\mathrm{T}\right)^\dag = \beta_l \mathbf{H}_l^*\left(\mathbf{H}_l^\mathrm{T} \mathbf{H}_l^* \right)^{-1},
\label{eq:zf}
\end{align}
where $\beta_l$ is the power normalization factor to meet the power constraints of the BS and $(\cdot)^\dag$ denotes the pseudo-inverse. Note that in this paper, full rank of $\mathbf{H}_l^\mathrm{T}$ is assumed, i.e., $\mathbf{H}_l^\mathrm{T} \mathbf{H}_l^*$ in (\ref{eq:zf}) is nonsingular hence invertible, which is the same assumption as \cite{Rusek_massive_MIMO_overview,Liang_HB,Ying_HB,Alkhateeb_HP,Sohrabi_HB}. This assumption is generally valid under favorable propagation \cite{Rusek_massive_MIMO_overview,Ngo_Favorable}, and can still be achieved by properly grouping UEs even if under non-favorable propagation. As mentioned in Section \ref{subsec:system}, in this paper, 
$||\mathbf{W}_l||^2_{\mathrm{F}} = 1/L_{\mathrm{used}}$ for the $L_{\mathrm{used}}$ used subcarriers. As a result, $\beta_l$ can be written as 
\begin{align}
\beta_l = \frac{1}{L_{\mathrm{used}} \mathrm{Tr} \left[ \left( \mathbf{H}_l^\mathrm{T} \mathbf{H}_l^* \right)^{-1} \right]}.
\label{eq:zf_normfactor}
\end{align}
Note that the ZF precoding requires full CSI $\mathbf{H}^\mathrm{T}_l$, which is very challenging or even impossible to be acquired in the considered case where $R\ll M$ as explained in Section I. 

As explained in Section \ref{sec:introduction}, previous works on HB \cite{Wu_HB,Ayach_mmWave,Alkhateeb_HB_CE,Geng_MUHB,Liang_HB,
Bogale_MUHB,Xu_HB,Chiu_HB,Ying_HB,Stirling_MUHB,Alkhateeb_HP,Sohrabi_HB,Sohrabi_HB_OFDM} are not well suited for the considered system model described in Section \ref{subsec:system}. 
Hence, an unified AB designed for all served groups of UEs is needed.
The goal of this section is to design a HB method for finding $\mathbf{W}_l = \mathbf{W}^{\mathrm{RF}} \mathbf{W}_l^{\mathrm{BB}}$ in (\ref{eq:precoding}) that can offer the performance close to the benchmark ZF precoding for massive MIMO systems. Let $\mathbf{U}$ be an $M\times M$ unitary matrix with each nonzero element having the same amplitude. Then, ZF precoding (\ref{eq:zf}) can be rewritten as
\begin{align}
\mathbf{W}^{\mathrm{ZF}}_l = \beta_l \mathbf{U} \mathbf{U}^\mathrm{H} \mathbf{H}_l^*\left(\mathbf{H}_l^\mathrm{T} \mathbf{U} \mathbf{U}^\mathrm{H} \mathbf{H}_l^* \right)^{-1}.
\label{eq:zf_alt}
\end{align}
The matrix $\mathbf{U}$ can be separated into two matrices as
\begin{align}
\mathbf{U} = \left[\mathbf{U}_{\mathrm{p}} \,\, \mathbf{U}_{\mathrm{n}} \right],
\label{eq:unitary}
\end{align}
where $\mathbf{U}_{\mathrm{p}}$ and $\mathbf{U}_{\mathrm{n}}$ are $M\times R$ and $M\times (M-R)$ matrices respectively. Then, $\mathbf{H}^\mathrm{T}_l \mathbf{U} 
= [\mathbf{H}_{\mathrm{p},l}^\mathrm{T} \,\, \mathbf{H}_{\mathrm{n},l}^\mathrm{T} ]$ 
where $\mathbf{H}_{\mathrm{p},l}^\mathrm{T} = \mathbf{H}^\mathrm{T}_l\mathbf{U}_{\mathrm{p}}$ and $\mathbf{H}_{\mathrm{n},l}^\mathrm{T} = \mathbf{H}^\mathrm{T}_l\mathbf{U}_{\mathrm{n}}$ are two $K_l\times R$ and $K_l\times (M-R)$ matrices that are derived by projecting $\mathbf{H}_l^\mathrm{T}$ into two subspaces spanned by $\mathbf{U}_{\mathrm{p}}$ and $\mathbf{U}_{\mathrm{n}}$ respectively. 
If $\mathbf{U}_{\mathrm{n}}$ is a null space of $\mathbf{H}_l^\mathrm{T}$, i.e., 
\begin{align}
\mathbf{H}^{\mathrm{T}}_{\mathrm{n},l} = 
\mathbf{H}^\mathrm{T}_l\mathbf{U}_{\mathrm{n}} = \mathbf{0}_{K_l\times \left(M-R\right)},
\label{eq:null_subcarrier}
\end{align}
where $\mathbf{0}_{X\times Y}$ denotes a $X\times Y$ all-zero matrix, which indicates that $\mathbf{H}_l^\mathrm{T}$ is located in the subspace of $\mathbf{U}_{\mathrm{p}}$ and $\mathbf{H}_{\mathrm{p},l}^\mathrm{T} = \mathbf{H}^\mathrm{T}_l\mathbf{U}_{\mathrm{p}}$ is full rank, then (\ref{eq:zf_alt}) can be rewritten as
\begin{align}
\mathbf{W}^{\mathrm{ZF}}_l = 
\beta_l \mathbf{U}_{\mathrm{p}} \mathbf{H}_{\mathrm{p},l}^* \left(\mathbf{H}_{\mathrm{p},l}^\mathrm{T} \mathbf{H}_{\mathrm{p},l}^* \right)^{-1}.
\label{eq:zf_hybrid}
\end{align}
Based on (\ref{eq:zf_hybrid}), the desired HB method can be constructed, where the baseband digital precoding matrix $\mathbf{W}_l^{\mathrm{BB}}$ in (\ref{eq:precoding}) for the $l$th subcarrier can apply ZF precoding as 
\ifCLASSOPTIONonecolumn
\begin{align}
\mathbf{W}^{\mathrm{BB}}_l = \beta_l \mathbf{H}_{\mathrm{p},l}^* \left(\mathbf{H}_{\mathrm{p},l}^\mathrm{T} \mathbf{H}_{\mathrm{p},l}^* \right)^{-1} = \beta_l \left(\mathbf{H}_{\mathrm{p},l}^\mathrm{T}\right)^\dag = 
\beta_l \mathbf{U}^{\mathrm{H}}_\mathrm{p} \mathbf{H}_l^* \left(\mathbf{H}^\mathrm{T}_l\mathbf{U}_{\mathrm{p}} \mathbf{U}^{\mathrm{H}}_\mathrm{p} \mathbf{H}_l^* \right)^{-1},
\label{eq:zf_digital}
\end{align}
\else
\begin{align}
\mathbf{W}^{\mathrm{BB}}_l = \beta_l \mathbf{H}_{\mathrm{p},l}^* \left(\mathbf{H}_{\mathrm{p},l}^\mathrm{T} \mathbf{H}_{\mathrm{p},l}^* \right)^{-1}
,
\label{eq:zf_digital}
\end{align}
\fi
while the RF analog precoding matrix $\mathbf{W}^{\mathrm{RF}}$ in (\ref{eq:precoding}) can be constructed as
\begin{align}
\mathbf{W}^{\mathrm{RF}} = \mathbf{U}_{\mathrm{p}}.
\label{eq:zf_analog}
\end{align}
Note that 
$\mathbf{H}_{\mathrm{p},l}^\mathrm{T} = \mathbf{H}^\mathrm{T}_l\mathbf{U}_{\mathrm{p}}$ (\ref{eq:zf_digital}) is the effective channel matrix seen at the baseband applying the AB constructed in (\ref{eq:zf_analog}), which can be estimated by conventional uplink training methods. With (\ref{eq:zf_digital}) and (\ref{eq:zf_analog}), the sum rate of the proposed HB with $R$ RF chains can be the same as the benchmark ZF precoding with $M$ RF chains. 

Based on (\ref{eq:unitary}), (\ref{eq:zf_digital}) and (\ref{eq:zf_analog}), the design goal becomes to constructing the unitary matrix $\mathbf{U}$ where the condition (\ref{eq:null_subcarrier}) is satisfied for each subcarrier. According to (\ref{eq:channel_freq}), the frequency-domain channel vector for a UE at any subcarrier is located in the same spatial subspace constructed by $\mathbf{a}(\phi_{ijk},\theta_{ijk})$, i.e., $\mathbf{V}_k$ in (\ref{eq:subspace}). Hence, the condition (\ref{eq:null_subcarrier}) can be replaced by 
\begin{align}
\mathbf{V}^\mathrm{T}_k \mathbf{U}_{\mathrm{n}} = \mathbf{0}_{N_{\mathrm{cl},k}N_{\mathrm{ray},k} \times \left(M-R\right)}, \,\, \forall k.
\label{eq:null}
\end{align}
Note that if (\ref{eq:null}) is not strictly hold, assuming that $\mathbf{H}_{\mathrm{p},l}^\mathrm{T} = \mathbf{H}^\mathrm{T}_l\mathbf{U}_{\mathrm{p}}$ is still full rank, which can be achieved by properly grouping UEs, the power normalization factor $\beta_l$ of (\ref{eq:zf_normfactor}) in (\ref{eq:zf_digital}) becomes
\begin{align}
\beta_{\mathrm{p},l} = \frac{1}{L_{\mathrm{used}} \mathrm{Tr} \left[ \left( \mathbf{H}_{\mathrm{p},l}^\mathrm{T} \mathbf{H}_{\mathrm{p},l}^* \right)^{-1} \right]}.
\label{eq:zf_digital_normfactor}
\end{align}
\ifCLASSOPTIONonecolumn
Note that $\mathbf{H}_l^\mathrm{T} \mathbf{H}_l^* = \mathbf{H}_l^\mathrm{T} \mathbf{U} \mathbf{U}^{\mathrm{H}} \mathbf{H}_l^* = [\mathbf{H}_{\mathrm{p},l}^\mathrm{T} \,\, \mathbf{H}_{\mathrm{n},l}^\mathrm{T} ] [\mathbf{H}_{\mathrm{p},l}^* \,\, \mathbf{H}_{\mathrm{n},l}^* ]^\mathrm{T} = \mathbf{H}_{\mathrm{p},l}^\mathrm{T}\mathbf{H}_{\mathrm{p},l}^* + \mathbf{H}_{\mathrm{n},l}^\mathrm{T} \mathbf{H}_{\mathrm{n},l}^* $. According to \cite{Horn_Matrix}, 
$\mathbf{H}_l^\mathrm{T} \mathbf{H}_l^* \succ 0$, $\mathbf{H}_{\mathrm{p},l}^\mathrm{T}\mathbf{H}_{\mathrm{p},l}^* \succ 0$, and $\mathbf{H}_{\mathrm{n},l}^\mathrm{T} \mathbf{H}_{\mathrm{n},l}^*\succeq 0$. Since 
$\mathbf{H}_l^\mathrm{T} \mathbf{H}_l^* \succeq \mathbf{H}_{\mathrm{p},l}^\mathrm{T}\mathbf{H}_{\mathrm{p},l}^*$, 
then $(\mathbf{H}_l^\mathrm{T} \mathbf{H}_l^*)^{-1} \preceq (\mathbf{H}_{\mathrm{p},l}^\mathrm{T}\mathbf{H}_{\mathrm{p},l}^*)^{-1}$ \cite{Horn_Matrix}. Hence, $\mathrm{Tr} [ ( \mathbf{H}_l^\mathrm{T} \mathbf{H}_l^* )^{-1}] \preceq \mathrm{Tr} [ ( \mathbf{H}_{\mathrm{p},l}^\mathrm{T} \mathbf{H}_{\mathrm{p},l}^* )^{-1}]$ according to \cite{Horn_Matrix}. Therefore, $\beta_{\mathrm{p},l} \leq \beta_l$, with equality if and only if (\ref{eq:null_subcarrier}) is satisfied, which implies that if (\ref{eq:null}) is not satisfied, performance losses at some subcarriers occur. Unfortunately, in practice, it might be hard to find $\mathbf{U}_{\mathrm{n}}$ that strictly satisfies (\ref{eq:null}) for all $K$ UEs associated to the BS since $K$ could be very large where $K_l\leq K \leq \sum_{l=1}^L K_l$. Hence, (\ref{eq:null}) is practically relaxed as
\else
Note that $\mathbf{H}_l^\mathrm{T} \mathbf{H}_l^* = \mathbf{H}_l^\mathrm{T} \mathbf{U} \mathbf{U}^{\mathrm{H}} \mathbf{H}_l^* = [\mathbf{H}_{\mathrm{p},l}^\mathrm{T} \,\, \mathbf{H}_{\mathrm{n},l}^\mathrm{T} ] [\mathbf{H}_{\mathrm{p},l}^* \,\, \mathbf{H}_{\mathrm{n},l}^* ]^\mathrm{T} $ $= \mathbf{H}_{\mathrm{p},l}^\mathrm{T}\mathbf{H}_{\mathrm{p},l}^* + \mathbf{H}_{\mathrm{n},l}^\mathrm{T} \mathbf{H}_{\mathrm{n},l}^* $. According to \cite{Horn_Matrix}, 
$\mathbf{H}_l^\mathrm{T} \mathbf{H}_l^* \succ 0$, $\mathbf{H}_{\mathrm{p},l}^\mathrm{T}\mathbf{H}_{\mathrm{p},l}^* \succ 0$, and $\mathbf{H}_{\mathrm{n},l}^\mathrm{T} \mathbf{H}_{\mathrm{n},l}^*\succeq 0$. Due to the fact that 
$\mathbf{H}_l^\mathrm{T} \mathbf{H}_l^* \succeq \mathbf{H}_{\mathrm{p},l}^\mathrm{T}\mathbf{H}_{\mathrm{p},l}^*$, 
then $(\mathbf{H}_l^\mathrm{T} \mathbf{H}_l^*)^{-1} \preceq (\mathbf{H}_{\mathrm{p},l}^\mathrm{T}\mathbf{H}_{\mathrm{p},l}^*)^{-1}$ based on \cite{Horn_Matrix}. Hence, $\mathrm{Tr} [ ( \mathbf{H}_l^\mathrm{T} \mathbf{H}_l^* )^{-1}] \preceq \mathrm{Tr} [ ( \mathbf{H}_{\mathrm{p},l}^\mathrm{T} \mathbf{H}_{\mathrm{p},l}^* )^{-1}]$ according to \cite{Horn_Matrix}. Therefore, $\beta_{\mathrm{p},l} \leq \beta_l$, with equality if and only if (\ref{eq:null_subcarrier}) is satisfied, which implies that if (\ref{eq:null}) is not satisfied, performance losses at some subcarriers occur. Unfortunately, in practice, it might be hard to find $\mathbf{U}_{\mathrm{n}}$ that strictly satisfies (\ref{eq:null}) for all $K$ UEs associated to the BS since $K$ could be very large where $K_l\leq K \leq \sum_{l=1}^L K_l$. Hence, (\ref{eq:null}) is practically relaxed as
\fi
\begin{align}
\min \sum_{l=1}^L \sum_{k_l=1}^{K_l}\|\mathbf{V}^\mathrm{T}_{k_l} \mathbf{U}_{\mathrm{n}}\|_\mathrm{F}^2,
\label{eq:null_relax}
\end{align}
where $\mathbf{V}^\mathrm{T}_{k_l}$ denotes the spanned subspace for the $k_l$th UE scheduled at the $l$th subcarrier. Note that the condition (\ref{eq:null_relax}) is equivalent to
\begin{align}
\max \sum_{l=1}^L \sum_{k=1}^{K_l}\|\mathbf{V}^\mathrm{T}_{k_l} \mathbf{U}_{\mathrm{p}}\|_\mathrm{F}^2 
& = \max \mathrm{tr} \left[ \left(\mathbf{W}^{\mathrm{RF}}\right)^{\mathrm{H}} \mathbf{R} \mathbf{W}^{\mathrm{RF}} \right],
\label{eq:max_proj}
\end{align}
where $\mathbf{R} = \sum_{l=1}^L \sum_{k_l=1}^{K_l} \mathbf{R}_{k_l}$ with $\mathbf{R}_{k_l} = \mathbf{V}_{k_l}^\mathrm{*} \mathbf{V}^\mathrm{T}_{k_l}$ being the $M\times M$ SCM for the $k_l$th UE scheduled at the $l$th subcarrier. Note that $\mathbf{R}_{k_l}$ is the same for all subcarriers. If all $M$ antennas are assumed to be connected to all $R$ RF chains, i.e., with global connection, the design problem can be formulated as 
\begin{align}
& \max \mathrm{tr} \left[ \left(\mathbf{W}^{\mathrm{RF}}\right)^{\mathrm{H}} \mathbf{R} \mathbf{W}^{\mathrm{RF}} \right], \nonumber \\
& \,\, \mathrm{s.t.} \,\, \left(\mathbf{W}^{\mathrm{RF}}\right)^{\mathrm{H}} \mathbf{W}^{\mathrm{RF}} = \mathbf{I}_R, \nonumber \\ & \qquad \left| w^{\mathrm{RF}}_{mr} \right| = \frac{1}{\sqrt{M}}.
\label{eq:design_problem}
\end{align}
Note that (\ref{eq:design_problem}) only requires the knowledge of SCM instead
of full CSI, which is a much more practical assumption than previous HB works as explained in Section \ref{sec:introduction}. If all $R$ RF chains are not connected to all $M$ antennas, i.e., with partial connection, (\ref{eq:design_problem}) only needs to change the constraints $| w^{\mathrm{RF}}_{mr} | = 1/\sqrt{M}$ to the associated partially connected PSN. Note that in this paper, 
global connection 
is focused, which is the same assumption as in \cite{Ayach_mmWave,Alkhateeb_HB_CE,Liang_HB,
Bogale_MUHB,Ying_HB,Alkhateeb_HP,Sohrabi_HB_OFDM,Sohrabi_HB}
, to derive a unified AB matrix $\mathbf{W}^{\mathrm{RF}}$ with upper-bound performance. The more practical partial connection case is considered in future work.  

\subsection{Design Problem Validation} \label{subsec:validation}

According to 
(\ref{eq:io_bb})
, the maximum achievable sum rate for the $l$th subcarrier can be realized by DPC based on $(\mathbf{H}_l^{\mathrm{BB}})^{\mathrm{T}}$ 
\cite{Caire_MIMO_Broadcast,Viswanath_Broadcast}. In \cite{Caire_MIMO_Broadcast}, it proves that at the high Signal-to-Noise Ratio (SNR) region, MU-MIMO capacity with single-antenna UEs can be approached by ZF Tomlinson-Harashima Precoding (ZF-THP) based on QR decomposition. Specifically, applying QR decomposition to $(\mathbf{H}_l^{\mathrm{BB}})^{\mathrm{T}}$ as $(\mathbf{H}_l^{\mathrm{BB}})^{\mathrm{T}} = \mathbf{H}_l^{\mathrm{T}}\mathbf{W}^{\mathrm{RF}} = \mathbf{G}\mathbf{Q}$, 
where $\mathbf{G}$ is a $K_l\times K_l$ upper-triangular matrix, and $\mathbf{Q}$ is a $K_l \times R$ matrix satisfying $\mathbf{Q}\mathbf{Q}^\mathrm{H} = \mathbf{I}_{K_l}$, the achievable sum rate for the $l$th subcarrier at the high SNR region is written as  
\begin{align}
C_l 
\approx \mathrm{E} \left[ \sum_{k_l=1}^{K_l} \log_2 \left(\xi_{l} g^2_{k_l k_l} \right) \right] 
= \chi_{l} + \mathrm{E} \left\lbrace \log_2 \det \left[ \mathbf{B}_l \right] \right\rbrace,
\label{eq:rate_subcarrier}
\end{align}
where $\xi_{l}$ is the solution of the water-filling power allocation \cite{Caire_MIMO_Broadcast,Goldsmith_WC_Ch10}, $\chi_{l} = \mathrm{E} \left[\sum_{k_l=1}^{K_l} \log_2 \xi_{l}\right]$, the $K_l \times K_l$ matrix $\mathbf{B}_l$ is $\mathbf{B}_l = \mathbf{H}_l^{\mathrm{T}}\mathbf{W}^{\mathrm{RF}}(\mathbf{W}^{\mathrm{RF}})^\mathrm{H} \mathbf{H}_l^{\mathrm{*}}$,
and $g_{k_l k_l}$ is the $k_l$th diagonal element of $\mathbf{G}$. Since $\mathbf{B}_l$ is positive definite, $\log_2 \det [\mathbf{B}_l]$ is concave. Then, based on Jensen's inequality \cite{Rudin_Analysis}, 
\begin{align}
C_l \leq  \chi_{l} +  \log_2 \det \mathrm{E} \left[ \mathbf{B}_l \right].
\label{eq:rate_subcarrier_jensen}
\end{align}
\ifCLASSOPTIONonecolumn
Based on the Karhunen-Loeve representation \cite{Adhikary_JSDM,Jiang_FDD_Massive_MIMO}, the channel vector $\mathbf{h}_{k_ll}$ can be written as $\mathbf{h}_{k_ll} = \mathbf{R}_{k_l}^{\frac{1}{2}} \mathbf{z}_{k_ll}$,
where $\mathbf{z}_{k_ll}$ is a $K_l$-dimensional vector with each element being a zero-mean unit-variance complex Gaussian random variable. Then,
\begin{align}
\mathrm{E} \left[b_{ijl}\right] & = \mathrm{E} \left[ \mathbf{z}_{\left(k_l=i\right)l}^\mathrm{T} \left(\mathbf{R}_{k_l=i}^{\frac{1}{2}}\right)^\mathrm{T} \mathbf{W}^{\mathrm{RF}}\left(\mathbf{W}^{\mathrm{RF}}\right)^\mathrm{H} \left(\mathbf{R}_{k_l=j}^{\frac{1}{2}}\right)^* \mathbf{z}_{\left(k_l=j\right)l}^* \right] \nonumber \\
& = \mathrm{tr}\left\lbrace \mathrm{E} \left[ \mathbf{z}_{\left(k_l=j\right)l}^* \mathbf{z}_{\left(k_l=i\right)l}^\mathrm{T} \right] \left(\mathbf{R}_{k_l=i}^{\frac{1}{2}}\right)^\mathrm{T} \mathbf{W}^{\mathrm{RF}}\left(\mathbf{W}^{\mathrm{RF}}\right)^\mathrm{H}\left(\mathbf{R}_{k_l=j}^{\frac{1}{2}}\right)^* \right\rbrace \nonumber \\ 
& = \left\lbrace 
\begin{array}{cc}
 \mathrm{tr}\left[ \left(\mathbf{R}_{k_l=i}^{\frac{1}{2}}\right)^\mathrm{T} \mathbf{W}^{\mathrm{RF}}\left(\mathbf{W}^{\mathrm{RF}}\right)^\mathrm{H}\left(\mathbf{R}_{k_l=j}^{\frac{1}{2}}\right)^* \right], & i=j,\\
0& i\neq j,
\end{array}
 \right. 
\label{eq:new_matrix_element}
\end{align} 
\else
The channel vector $\mathbf{h}_{k_ll}$ can be written as $\mathbf{h}_{k_ll} = \mathbf{R}_{k_l}^{\frac{1}{2}} \mathbf{z}_{k_ll}$ by the Karhunen-Loeve representation \cite{Adhikary_JSDM,Jiang_FDD_Massive_MIMO},
where $\mathbf{z}_{k_ll}$ is a $K_l$-dimensional vector with each element being a zero-mean unit-variance complex Gaussian random variable. Then,
\begin{align}
\mathrm{E} \left[b_{ijl}\right] & = \mathrm{E} \left[ \mathbf{z}_{il}^\mathrm{T} \left(\mathbf{R}_{i}^{\frac{1}{2}}\right)^\mathrm{T} \mathbf{W}^{\mathrm{RF}}\left(\mathbf{W}^{\mathrm{RF}}\right)^\mathrm{H} \left(\mathbf{R}_{j}^{\frac{1}{2}}\right)^* \mathbf{z}_{jl}^* \right] \nonumber \\
& = \mathrm{tr}\left\lbrace \mathrm{E} \left[ \mathbf{z}_{jl}^* \mathbf{z}_{il}^\mathrm{T} \right] \left(\mathbf{R}_{i}^{\frac{1}{2}}\right)^\mathrm{T} \mathbf{W}^{\mathrm{RF}}\left(\mathbf{W}^{\mathrm{RF}}\right)^\mathrm{H}\left(\mathbf{R}_{j}^{\frac{1}{2}}\right)^* \right\rbrace \nonumber \\ 
& = \left\lbrace 
\begin{array}{cc}
 \mathrm{tr}\left[ \left(\mathbf{R}_{i}^{\frac{1}{2}}\right)^\mathrm{T} \mathbf{W}^{\mathrm{RF}}\left(\mathbf{W}^{\mathrm{RF}}\right)^\mathrm{H}\left(\mathbf{R}_{j}^{\frac{1}{2}}\right)^* \right], & i=j,\\
0& i\neq j,
\end{array}
 \right. 
\label{eq:new_matrix_element}
\end{align} 
\fi
where $b_{ijl}$ is the $\{i,j\}$th element in $\mathbf{B}_l$. This equation indicates that $\mathrm{E}[\mathbf{B}_l]$ is a diagonal matrix. Then, $C_l$ is further written as 
\begin{align}
C_l 
\leq 
\chi_{l} +   \sum_{k_l=1}^{K_l}  \log_2 \mathrm{tr}\left[  \left(\mathbf{W}^{\mathrm{RF}}\right)^\mathrm{H} \mathbf{R}_{k_l} \mathbf{W}^{\mathrm{RF}} \right]. 
\label{eq:rate_subcarrier_cov}
\end{align}
Since $\log_2(x)$ is a concave function, based on Jensen's inequality \cite{Rudin_Analysis}, 
\begin{align}
C_l 
\leq \chi_{l} + K_l \log_2 \mathrm{tr}\left[  \left(\mathbf{W}^{\mathrm{RF}}\right)^\mathrm{H} \left(\frac{\sum_{k_l=1}^{K_l} \mathbf{R}_{k_l}}{K_l}   \right) \mathbf{W}^{\mathrm{RF}} \right].
\label{eq:rate_subcarrier_final}
\end{align} 
Hence, applying Jensen's inequality again, the system achievable sum rate is upper bounded as
\ifCLASSOPTIONonecolumn
\begin{align}
C 
 = \sum_{l=1}^L C_l 
\leq \sum_{l=1}^L \chi_{l} + \left(\sum_{l=1}^L K_l\right) \log_2 \mathrm{tr}\left[  \left(\mathbf{W}^{\mathrm{RF}}\right)^\mathrm{H} \left(\frac{\sum_{l=1}^L \sum_{k_l=1}^{K_l} \mathbf{R}_{k_l}}{\sum_{l=1}^L K_l}  \right) \mathbf{W}^{\mathrm{RF}} \right] 
. 
\label{eq:rate}
\end{align} 
Based on (\ref{eq:rate}), maximizing the upper bound of $C$ equals to the same design problem of (\ref{eq:design_problem})
.
\else
\begin{align}
C 
 = \sum_{l=1}^L C_l 
& \leq \sum_{l=1}^L \chi_{l} + K' \log_2 \mathrm{tr}\left[  \left(\mathbf{W}^{\mathrm{RF}}\right)^\mathrm{H} \left(\frac{\mathbf{R}}{K'}  \right) \mathbf{W}^{\mathrm{RF}} \right] 
,
\label{eq:rate}
\end{align} 
\fi
where $K' = \sum_{l=1}^{L} K_l$. Based on (\ref{eq:rate}), maximizing the upper bound of $C$ equals to the same design problem of (\ref{eq:design_problem})
.
  
\subsection{Unified Analog Beamforming} \label{subsec:analog}
The design problem (\ref{eq:design_problem}) derived in Section \ref{subsec:problem} and further validated in Section \ref{subsec:validation} is an optimization problem with feasible region \cite{Boyd_CO} in Grassmann manifold \cite{Conway_Grassmann}, which involves extremely complicated iterations as $M$ is very large. As a result, the design problem it is not well suited for practical massive MIMO-OFDM systems. Therefore, an explicit suboptimal solution is derived in this subsection
, assuming the knowledge of SCM $\mathbf{R}_{k_l}$ for each UE.

First, applying eigenvalue decomposition \cite{Adhikary_JSDM,Jiang_FDD_Massive_MIMO} to $\mathbf{R}$ as
\begin{align}
\mathbf{R} = \sum_{l=1}^L \sum_{k=1}^{K_l} \mathbf{R}_{k_l} = \mathbf{P} \mathbf{M} \mathbf{P}^\mathrm{H},
\label{eq:eigen}
\end{align}
where $\mathbf{P}$ is an $M\times M$ unitary matrix, and $\mathbf{M}$ is an $M\times M$ diagonal matrix whose diagonal elements $\mu_m$ are eigenvalues of $\mathbf{R}$ in decreasing order, then the objective function of (\ref{eq:design_problem}) can be rewritten as
\ifCLASSOPTIONonecolumn
\begin{align}
\max \mathrm{tr} \left[ \left(\mathbf{W}^{\mathrm{RF}}\right)^{\mathrm{H}} \mathbf{R} \mathbf{W}^{\mathrm{RF}} \right] = \max \mathrm{tr} \left[ \left(\mathbf{W}^{\mathrm{RF}}\right)^{\mathrm{H}} \mathbf{P} \mathbf{M} \mathbf{P}^\mathrm{H} \mathbf{W}^{\mathrm{RF}} \right] = \max \sum_{m=1}^M c_m \mu_m,
\label{eq:obj_new}
\end{align}
\else
\begin{align}
\max \mathrm{tr} \left[ \left(\mathbf{W}^{\mathrm{RF}}\right)^{\mathrm{H}} \mathbf{P} \mathbf{M} \mathbf{P}^\mathrm{H} \mathbf{W}^{\mathrm{RF}} \right] = \max \sum_{m=1}^M c_m \mu_m,
\label{eq:obj_new}
\end{align}
\fi
where
\begin{align}
0 \leq c_m = \| \left(\mathbf{W}^{\mathrm{RF}}\right)^{\mathrm{H}} \mathbf{p}_m\|^2_2 = \|\mathbf{U}_\mathrm{p}^{\mathrm{H}} \mathbf{p}_m\|^2_2 \leq \|\mathbf{U}^\mathrm{H} \mathbf{p}_m\|^2_2 = 1,
\label{eq:factor_eigen}
\end{align}
with $\mathbf{p}_m$ being the $m$th column of $\mathbf{P}$. Note that
\ifCLASSOPTIONonecolumn
\begin{align}
\sum_{m=1}^M c_m = \sum_{m=1}^M \| \left(\mathbf{W}^{\mathrm{RF}}\right)^{\mathrm{H}} \mathbf{p}_m\|^2_2 = \|\left(\mathbf{W}^{\mathrm{RF}}\right)^{\mathrm{H}} \mathbf{P}\|^2_\mathrm{F} = \mathrm{tr} \left\lbrace \mathbf{I}_R \right\rbrace = R.
\label{eq:factor_eigen_sum}
\end{align}
\else
\begin{align}
\sum_{m=1}^M c_m 
= \|\left(\mathbf{W}^{\mathrm{RF}}\right)^{\mathrm{H}} \mathbf{P}\|^2_\mathrm{F} = \mathrm{tr} \left\lbrace \mathbf{I}_R \right\rbrace = R.
\label{eq:factor_eigen_sum}
\end{align}
\fi
Based on (\ref{eq:factor_eigen}) and (\ref{eq:factor_eigen_sum}), the objection function (\ref{eq:obj_new}) becomes 
\begin{align}
\max \sum_{m=1}^M c_m \mu_m = \sum_{r=1}^R \mu_r.
\label{eq:obj_final}
\end{align}
The maximum value of $\sum_{r=1}^R \mu_r$ in (\ref{eq:obj_final}) can be achieved by 
\begin{align}
\mathbf{W}^{\mathrm{RF}} = \left[\mathbf{p}_i \right]_{i=1}^R =  \mathbf{P}_\mathrm{p}.
\label{eq:solution}
\end{align}
Note that if the AB can adjust both phases and amplitudes, then (\ref{eq:solution}) is the unified AB matrix. Unfortunately, the solution (\ref{eq:solution}) generally does not satisfy the constrains $| w^{\mathrm{RF}}_{mr} | = 1/\sqrt{M}$ in (\ref{eq:design_problem}) for the globally connected PSN, which cannot be relaxed. Hence, the constraints $(\mathbf{W}^{\mathrm{RF}})^{\mathrm{H}} \mathbf{W}^{\mathrm{RF}} = \mathbf{I}_R$ in (\ref{eq:design_problem}) are relaxed as $||\mathbf{W}^{\mathrm{RF}} - \mathbf{P}_\mathrm{p}||_\mathrm{F}^2 \approx 0$. Then, the relaxed design problem is 
\begin{align}
& \min \|\mathbf{W}^{\mathrm{RF}} - \mathbf{P_\mathrm{p}}\|_\mathrm{F}^2 \nonumber \\
& \,\, \mathrm{s.t.} \left| w^{\mathrm{RF}}_{mr} \right| = \frac{1}{\sqrt{M}}.
\label{eq:design_problem_relax}
\end{align}
Note that 
\ifCLASSOPTIONonecolumn
\begin{align}
\|\mathbf{W}^{\mathrm{RF}} - \mathbf{P}_\mathrm{p}\|_\mathrm{F}^2 = \|\mathbf{W}^{\mathrm{RF}}\|_\mathrm{F}^2 + \|\mathbf{P}_\mathrm{p}\|_\mathrm{F}^2 - 2\mathrm{Tr}\left\lbrace \Re \left[ \mathbf{W}^{\mathrm{RF}}\mathbf{P}_\mathrm{p}^\mathrm{H} \right] \right\rbrace = 2R - 2\mathrm{Tr}\left\lbrace \Re\left[ \mathbf{W}^{\mathrm{RF}} \mathbf{P}_\mathrm{p}^\mathrm{H} \right] \right\rbrace,
\label{eq:obj_relax}
\end{align}
\else
\begin{align}
\|\mathbf{W}^{\mathrm{RF}} - \mathbf{P}_\mathrm{p}\|_\mathrm{F}^2 = 2R - 2\mathrm{Tr}\left\lbrace \Re\left[ \mathbf{W}^{\mathrm{RF}} \mathbf{P}_\mathrm{p}^\mathrm{H} \right] \right\rbrace,
\label{eq:obj_relax}
\end{align}
\fi
where $\Re[\cdot]$ denotes the real part of the input value, whose minimum value is achieved when the $M\times R$ matrix $\mathbf{W}^{\mathrm{RF}}$ has same phase components as $\mathbf{P}_\mathrm{p}$, i.e.,
\begin{align}
\mathbf{W}^{\mathrm{RF}} = \frac{1}{\sqrt{M}} \exp\left\lbrace j \arg \left(\mathbf{P}_\mathrm{p} \right) \right\rbrace,
\label{eq:solution_relax}
\end{align}
where $\arg(\cdot)$ denotes the argument operator. The solution (\ref{eq:solution_relax}) is the proposed unified AB matrix according to (\ref{eq:eigen}) and (\ref{eq:solution})
. Note that although the solution (\ref{eq:solution_relax}) is derived for TDD massive MIMO-OFDM systems in sub $6\mathrm{GHz}$ bands, as long as the SCM is available, it can be easily applied to FDD systems and single-carrier SU-MIMO or MU-MIMO systems. In addition, it can be applied to both LoS and NLoS channels. With simple generalization to the case of multiple-antenna UEs, it promises to be applicable to mmWave systems, which is better dealt in a separate paper, hence is not discussed in detail in this paper. 

\subsection{Number of RF Chains} \label{subsec:rf_chains}
Based on (\ref{eq:obj_final}), a larger $R$ generally results in an increased value of 
(\ref{eq:obj_new}) and the maximum value is surely achieved when $R=M$. Obviously, a trade-off exists between the performance of the unified AB proposed in Section \ref{subsec:analog} and the hardware cost that is mainly affected by the number of RF chains $R$. 

\ifCLASSOPTIONonecolumn
In the best case, if $R \geq \mathrm{rank}(\mathbf{R})$,
then (\ref{eq:obj_final}) reaches the maximum value since the eigenvalues $\mu_{R+1},\ldots,\mu_{M}$ all equal to zero. Based on \cite{Yin_CE_Massive_MIMO}, with supercritical antenna spacing, i.e., no more than half wavelength, when $M_\mathrm{h}$ and $M_\mathrm{v}$ are relatively large, the ranks of $\mathrm{span}\{\mathbf{a}_\mathrm{h}(\phi,\theta)\}$ and $\mathrm{span}\{ \mathbf{a}_\mathrm{v}(\theta) \}$ in (\ref{eq:upa}) 
can be approximated respectively as
\begin{align}
\kappa_\mathrm{h} = \mathrm{rank}\left\lbrace \mathrm{span}\left[ \mathbf{a}_\mathrm{h}\left(\phi,\theta \right) \right]  \right\rbrace \approx \lceil M_\mathrm{h} \frac{d_\mathrm{h}}{\lambda} \left(\max\left[\cos\phi \sin \theta \right] - \min \left[\cos \phi \sin \theta \right] \right) \rceil,
\label{eq:rank_horizontal}
\end{align} 
and
\begin{align}
\kappa_\mathrm{v} = \mathrm{rank}\left\lbrace \mathrm{span} \left[ \mathbf{a}_\mathrm{v}\left(\theta \right) \right]  \right\rbrace \approx \lceil M_\mathrm{v} \frac{d_\mathrm{v}}{\lambda} \left(\cos\theta_\mathrm{min} -\cos \theta_\mathrm{max}\right) \rceil,
\label{eq:rank_vertical}
\end{align} 
\else 
In the best case, if $R \geq \mathrm{rank}(\mathbf{R})$,
then (\ref{eq:obj_final}) reaches the maximum value since the eigenvalues $\mu_{R+1},\ldots,\mu_{M}$ all equal to zero. Based on \cite{Yin_CE_Massive_MIMO}, with supercritical antenna spacing, i.e., no more than half wavelength, when $M_\mathrm{h}$ and $M_\mathrm{v}$ are relatively large, the ranks of $\mathrm{span}\{\mathbf{a}_\mathrm{h}(\phi,\theta)\}$ and $\mathrm{span}\{ \mathbf{a}_\mathrm{v}(\theta) \}$ in (\ref{eq:upa}), $\kappa_\mathrm{h}$ and $\kappa_\mathrm{v}$, 
can be approximated respectively as
\begin{align}
\kappa_\mathrm{h} \approx \lceil M_\mathrm{h} \frac{d_\mathrm{h}}{\lambda} \left(\max\left[\cos\phi \sin \theta \right] - \min \left[\cos \phi \sin \theta \right] \right) \rceil,
\label{eq:rank_horizontal}
\end{align} 
and
\begin{align}
\kappa_\mathrm{v} \approx \lceil M_\mathrm{v} \frac{d_\mathrm{v}}{\lambda} \left(\cos\theta_\mathrm{min} -\cos \theta_\mathrm{max}\right) \rceil,
\label{eq:rank_vertical}
\end{align} 
\fi
where $\lceil \cdot \rceil$ denotes the ceiling operation. As a result,
\begin{align}
\kappa = \mathrm{rank}\left\lbrace \mathrm{span} \left[ \mathbf{a}\left(\phi,\theta\right) \right]\right\rbrace = \kappa_\mathrm{h}\kappa_\mathrm{v} 
.
\label{eq:rank_subspace}
\end{align} 
Then, a sufficient number of $R$ to achieve the performance close to complete digital beamforming 
is
\begin{align}
R \geq \kappa \geq \mathrm{rank}\left(\mathbf{R}\right).
\label{eq:r_lb_final}
\end{align}

According to (\ref{eq:rank_horizontal})-(\ref{eq:r_lb_final}), some observations are listed below.
\begin{enumerate}
\item{Note that $(\max[\cos\phi \sin \theta] - \min[\cos \phi \sin \theta]) \in [0,2]$, $(\cos\theta_\mathrm{min} -\cos \theta_\mathrm{max})\in [0,2]$, $d_\mathrm{h}/\lambda \leq 0.5$, and $d_\mathrm{v}/\lambda \leq 0.5$. Then, $\kappa_\mathrm{h}$ and $\kappa_\mathrm{v}$ can be much smaller than $M_\mathrm{h}$ and $M_\mathrm{v}$ respectively, so $\kappa$ can be much smaller than $M$, which indicates that the hardware cost can be greatly reduced by HB.}
\item{The values of $(\max[\cos\phi \sin \theta] - \min[\cos \phi \sin \theta])$ and $(\cos\phi_\mathrm{min} -\cos \phi_\mathrm{max})$ reduce when the horizontal and vertical spread angles become smaller respectively, so $\kappa$ becomes smaller, which indicates that $R$ can be reduced for antenna arrays with higher directivity.} 
\item{
For ULA, (\ref{eq:rank_subspace}) degenerates to (\ref{eq:rank_horizontal}) with $M_\mathrm{h}=M$ for horizontal ULA or (\ref{eq:rank_vertical}) with $M_\mathrm{v}=M$ for vertical ULA. As discussed in the first observation, since both $d_\mathrm{h}/\lambda (\max[\cos\phi \sin \theta] - \min[\cos \phi \sin \theta])$ and $d_\mathrm{v}/\lambda (\cos\theta_\mathrm{min} -\cos \theta_\mathrm{max})$ can be much smaller than $1$, the number of $R$ for UPA can be much smaller than ULA, which indicates that UPA is better suited for the proposed HB than ULA.}
\end{enumerate}

\subsection{Complexity} \label{subsec:complexity_hb}
The complexity of the proposed unified AB in Section \ref{subsec:analog} is dominated by the eigenvalue decomposition conducted in (\ref{eq:eigen}). Since the $M\times M$ SCM $\mathbf{R}$ in (\ref{eq:eigen}) is symmetric, according to \cite{Nielsen_Matrix}, the complexity is generally $\mathcal{O}(M^3)$. Because 
the change of $\mathbf{R}$ is relatively slow
, the proposed unified AB matrix can be employed for a relatively long time before being updated. As a result, the complexity is in fact very low.  
 
\section{Subspace Construction with Partial CSI} \label{sec:subspace}
In Section \ref{sec:hybrid}, the knowledge of the 
SCM $\mathbf{R}_{k_l}$ is required for the proposed HB algorithm with unified AB. 
Compared to the full 
CSI $\mathbf{h}_{k_ll}$, the change of $\mathbf{R}_{k_l}$ is much slower, so it can be well tracked in conventional systems \cite{Adhikary_JSDM,Jiang_FDD_Massive_MIMO}. However, in the case of HB, due to the fact that $M\gg R$, it is still a challenging problem to efficiently acquire $\mathbf{R}_{k_l}$. Specifically, in the case of globally connected PSN, since at most $R$ antennas can measure the 
CSI between them and each UE in an uplink training session, the training overhead is relatively large to acquire full 
CSI in order to estimate $\mathbf{R}_{k_l}$. In addition, for the more practical partially connected PSN, it is not even possible to acquire full 
CSI. To tackle this issue, in this section, a novel SC method based on only partial CSI is proposed to estimate $\mathbf{R}_{k_l}$ for the proposed HB in a TDD massive MIMO-OFDM system to achieve the performance close to the case with perfect SCM. To the best of our knowledge, no other literature has discussed this specific issue for HB. 

\subsection{Partial CSI} \label{subsec:partial_csi}
In TDD systems, the partial instantaneous CSI can be acquired by standard uplink channel estimation methods such as the cyclic shift method based on Zadoff-Chu (ZC) sequences employed in LTE \cite{3GPP_TS_36.211}, and used for both the uplink and downlink due to the channel reciprocity. For the sake of simplicity, in this section, perfect partial CSI is assumed in Section \ref{subsec:partial_csi} to Section \ref{subsec:dimension} to illustrate the proposed method. In addition, in Section \ref{subsec:robustness}, channel estimation error is considered. Hence, the $P \times 1$ partial instantaneous CSI with $1\leq P \leq R$ for the $k_l$th UE at the $l$th subcarrier can be represented as 
\begin{align}
\bar{\mathbf{h}}_{k_ll} = \mathbf{T} \mathbf{h}_{k_ll},
\label{eq:partial_csi}
\end{align}
where the $P\times M$ matrix $\mathbf{T}$ with each element being either $1$ or $0$ denotes the partial connection matrix when the uplink channel estimation is conducted. The partial connection matrix  $\mathbf{T}$ needs to be carefully selected in order to acquire a proper $\bar{\mathbf{h}}_{k_ll}$ to estimate $\mathbf{R}_{k_l}$.

\subsection{Subspace Construction} \label{subsec:subspace}
For UPA, based on (\ref{eq:subspace}), the subspace of each UE is the Kronecker product of its horizontal and vertical subspaces. Since the horizontal and vertical spatial features can be separately tracked by only one row and one column of the UPA respectively, their subspaces can be separately constructed with the condition 
\begin{align}
R \geq M_\mathrm{h}+M_\mathrm{v}-1.
\label{eq:sc_condition}
\end{align}
In addition, $\bar{\mathbf{h}}_{k_ll}$ 
in (\ref{eq:partial_csi}) needs to include at least one row and one column of the antenna array, e.g., 
\begin{align}
\mathbf{T} = \left[
\begin{array}{cccc}
\mathbf{I}_{M_\mathrm{h}} & \mathbf{0}_{M_\mathrm{h}\times M_\mathrm{h}} & \cdots & \mathbf{0}_{M_\mathrm{h}\times M_\mathrm{h}} \\
\mathbf{0}_{\left(M_\mathrm{v}-1\right)\times M_\mathrm{h}} &\mathbf{Z}_{1} & \cdots & \mathbf{Z}_{M_\mathrm{v}-1}
\end{array} 
\right],
\label{eq:tmatrix_eg}
\end{align}
with $P = M_\mathrm{h}+M_\mathrm{v}-1$, where $\mathbf{Z}_i$ denotes an $(M_\mathrm{v}-1)\times M_\mathrm{h}$ matrix with all zero elements except the first element of the $i$th row $z_{i1}=1$. With (\ref{eq:tmatrix_eg}), the horizontal and vertical instantaneous CSI vectors can be constructed respectively as
\begin{align}
\mathbf{h}_{k_ll}^\mathrm{h} = \left[\bar{h}_{1k_ll} \,\, \cdots \,\, \bar{h}_{M_\mathrm{h}k_ll}  \right]^\mathrm{T},
\label{eq:csi_horizontal}
\end{align}
and
\begin{align}
\mathbf{h}_{k_ll}^\mathrm{v} = \left[\bar{h}_{1k_ll} \,\, \bar{h}_{\left(M_\mathrm{h}+1\right)k_ll} \,\, \cdots \,\, \bar{h}_{\left(M_\mathrm{h}+M_\mathrm{v}-1\right)k_ll}  \right]^\mathrm{T},
\label{eq:csi_vertical}
\end{align}  
where $\bar{h}_{pk_ll}$ is $p$th element of $\bar{\mathbf{h}}_{k_ll}$ with $p=1,\ldots,P$. According to (\ref{eq:csi_horizontal}) and (\ref{eq:csi_vertical}), the horizontal and vertical subspaces can be constructed separately. Since the proposed SC method can be applied to both the horizontal and vertical dimensions, only the horizontal dimension is discussed below to present the proposed SC method.

First, the $M_\mathrm{h} \times M_\mathrm{h}$ SCM for the horizontal dimension 
is calculated as
\begin{align}
\mathbf{R}_{k_l}^\mathrm{h} = \mathrm{E} \left[ \mathbf{h}_{{k_l}l}^\mathrm{h} \left( \mathbf{h}_{{k_l}l}^\mathrm{h} \right)^\mathrm{H} \right],
\label{eq:cov_horizontal}
\end{align}  
where the expectation can be realized by averaging estimated channel vectors for the related UE at multiple subcarriers
. Then, the dominant angle $\omega_{0k_l}^\mathrm{h} \in [0,2\pi)$ is searched as 
\begin{align}
\omega_{0k_l}^\mathrm{h} =  \mathop {\arg\max} \limits_{\omega \in \left[0,2\pi \right)} \mathbf{e}_{M_\mathrm{h}}^\mathrm{H} \left[\omega \right] \mathbf{R}_{k_l}^\mathrm{h} \mathbf{e}_{M_\mathrm{h}}\left[\omega \right],
\label{eq:dominant_search}
\end{align}  
where the steering vector is defined as
\begin{align}
\mathbf{e}_{M_\mathrm{h}} \left[\omega \right] = \frac{1}{\sqrt{M_\mathrm{h}}} \left[1 \,\, e^{j\omega}\,\, \cdots \,\, e^{j\left(M_\mathrm{h}-1\right)\omega} \right]^\mathrm{T}, \,\, \omega \in \left[0,2\pi\right).
\label{eq:vector_steering}
\end{align} 
Define an $N_{\mathrm{F}}^{\mathrm{h}} \times 1$ special vector $\boldsymbol{\rho}_{k_l}^\mathrm{h}$ where $N_{\mathrm{F}}^{\mathrm{h}}\geq M_\mathrm{h}$ is a configurable parameter as
\ifCLASSOPTIONonecolumn
\begin{align}
\boldsymbol{\rho}_{k_l}^\mathrm{h} = \left[ \frac{1}{2} \sum_{m=1}^{M_\mathrm{h}} r_{mmk_l}^\mathrm{h} \,\, \sum_{m=1}^{M_\mathrm{h}-1} r_{m\left(m+1\right)k_l}^\mathrm{h} \,\, \sum_{m=1}^{M_\mathrm{h}-2} r_{m\left(m+2\right)k_l}^\mathrm{h} \,\,\cdots \,\, r_{1M_\mathrm{h}k_l}^\mathrm{h} \,\, 0 \,\, \cdots \,\, 0 \right]^\mathrm{T},
\label{eq:vector_spectial}
\end{align} 
\else
\begin{align}
\boldsymbol{\rho}_{k_l}^\mathrm{h} = & \left[ \frac{1}{2} \sum_{m=1}^{M_\mathrm{h}} r_{mmk_l}^\mathrm{h} \,\, \sum_{m=1}^{M_\mathrm{h}-1} r_{m\left(m+1\right)k_l}^\mathrm{h} \,\, \right. \nonumber \\ 
& \left. \sum_{m=1}^{M_\mathrm{h}-2} r_{m\left(m+2\right)k_l}^\mathrm{h} \,\,\cdots \,\, r_{1M_\mathrm{h}k_l}^\mathrm{h} \,\, 0 \,\, \cdots \,\, 0 \right]^\mathrm{T},
\label{eq:vector_spectial}
\end{align} 
\fi
where $r_{ijk_l}^\mathrm{h}$ is the $\{i,j\}$th element of $\mathbf{R}_{k_l}^\mathrm{h}$, with $i,j = 1,\ldots M_\mathrm{h}$. Then, (\ref{eq:dominant_search}) can be replaced by 
\begin{align}
\omega_{0k_l}^\mathrm{h} = \left. \frac{2\pi \left(n_{0k_l}^\mathrm{h}-1 \right)}{N_\mathrm{F}^{\mathrm{h}}} \,\, \right|  \,\, n_{0k_l}^\mathrm{h} = \mathop {\arg\max} \limits_{n \in \left[1,N_\mathrm{F}^{\mathrm{h}}\right]} \Re \left[ \mathcal{F}\left\lbrace \boldsymbol{\rho}_{k_l}^\mathrm{h} \right\rbrace^{N_\mathrm{F}^{\mathrm{h}}}_n \right],
\label{eq:dominant_search_fft}
\end{align}  
where $\mathcal{F}\{\cdot\}_n^\mathrm{N_\mathrm{F}^{\mathrm{h}}}$ denotes the $n$th element of the $N_\mathrm{F}^{\mathrm{h}}$-point FFT operation
. Note that the derivations of (\ref{eq:dominant_search_fft}) are provided in Appendix. Based on the dominant angle $\omega_{0k_l}^\mathrm{h}$ derived by (\ref{eq:dominant_search_fft}), an $M_\mathrm{h} \times M_\mathrm{h}$ steering matrix $\mathbf{S}^{\mathrm{h}}_{k_l}$ is constructed as 
\begin{align}
\mathbf{s}^{\mathrm{h}}_{mk_l} = \mathbf{e}_{M_\mathrm{h}} \left[\omega_{0k_l}^\mathrm{h} + \frac{2\pi(m-1)}{M_\mathrm{h}} \right], \,\, m=1,\ldots,M_\mathrm{h},
\label{eq:sc_vector_steering}
\end{align} 
where $\mathbf{s}^{\mathrm{h}}_{mk_l}$ denotes the $m$th column of $\mathbf{S}^{\mathrm{h}}_{k_l}$. Note that $\mathbf{S}^{\mathrm{h}}_{k_l}$ is a unitary matrix which is constructed to track the $M_\mathrm{h}$ angles associated with the dominant angle $\omega_{0k_l}^\mathrm{h}$. Define an $M_\mathrm{h} \times M_\mathrm{h}$ matrix $\mathbf{D}_{k_l}^\mathrm{h}$ with the $m$th diagonal element being  
\begin{align}
d^{\mathrm{h}}_{mmk_l} = \left(\mathbf{s}^{\mathrm{h}}_{mk_l}\right)^\mathrm{H} \mathbf{R}_{k_l}^\mathrm{h} \mathbf{s}^{\mathrm{h}}_{mk_l}.
\label{eq:diagonal_steering}
\end{align} 
Note that if $N_{\mathrm{F}}^{\mathrm{h}}$ is a multiple of $M_{\mathrm{h}}$, the value of $d^{\mathrm{h}}_{mmk_l}$ in (\ref{eq:diagonal_steering}) is available after (\ref{eq:dominant_search_fft}). Then, the horizontal subspace can be constructed as 
\begin{align}
\hat{\mathbf{V}}^{\mathrm{h}}_{k_l} = \frac{1}{ \left\|\mathbf{D}_{k_l}^\mathrm{h} \right\|_\mathrm{F}
 }  \mathbf{S}^{\mathrm{h}}_{k_l} \mathbf{D}_{k_l}^\mathrm{h}.
\label{eq:sc_horizontal}
\end{align} 
With the constructed horizontal and vertical subspaces, the subspace $\mathbf{V}_{k_l}$ in (\ref{eq:subspace}) can be constructed as 
\begin{align}
\hat{\mathbf{V}}_{k_l} = \hat{\mathbf{V}}^{\mathrm{h}}_{k_l} \otimes \hat{\mathbf{V}}^{\mathrm{v}}_{k_l}.
\label{eq:sc_total}
\end{align} 
Finally, $\mathbf{R}_{k_l}$ can be estimated as 
\begin{align}
\hat{\mathbf{R}}_{k_l} = \hat{\mathbf{V}}_{k_l}^* \hat{\mathbf{V}}_{k_l}^\mathrm{T}.
\label{eq:cov_est}
\end{align} 

If the condition (\ref{eq:sc_condition}) is not satisfied, then at least one of the horizontal and vertical dimensions can acquire the partial instantaneous CSI from only a subset of one row or one column of the antenna array in an uplink training session. Still taking the horizontal dimension as the example, assume that only the instantaneous CSI of $1\leq M_\mathrm{h}' < M_\mathrm{h}$ antennas of a row is available. Then, 
to apply the SC method proposed in this subsection, $M_\mathrm{h}'$ conservative antennas in one row should be selected, e.g., the first $M_\mathrm{h}'$ antennas in the first row. Then, (\ref{eq:csi_horizontal}), (\ref{eq:cov_horizontal})-(\ref{eq:diagonal_steering}) can be applied by replacing $M_\mathrm{h}$ with $M_\mathrm{h}'$. Then, in (\ref{eq:sc_horizontal}), instead of using the $M_\mathrm{h}'\times M_\mathrm{h}'$ matrix $\mathbf{S}^{\mathrm{h}}_{k_l}$, it is reconstructed into a $M_\mathrm{h}\times M_\mathrm{h}'$ matrix replacing each column based on $\mathbf{e}_{M_\mathrm{h}'}[\omega]$ with $\mathbf{e}_{M_\mathrm{h}}[\omega]$.

Note that due to the channel reciprocity in TDD systems, the estimated SCM can be applied to both the downlink and uplink. In addition, the SC method proposed in this subsection can also be applied to a proper partially connected PSN. In other words, a partially connected PSN should be designed to be capable of applying the proposed SC method. 

\subsection{Dimension Reduction} \label{subsec:dimension}
In fact, in (\ref{eq:sc_horizontal}), the size of the constructed horizontal subspace is $M_\mathrm{h}\times M_\mathrm{h}$, which may be further reduced to relax the computational complexity. Specifically, after (\ref{eq:diagonal_steering}), a threshold $\gamma$, e.g., $\gamma=100$, can be employed so that the constructed 
$\mathbf{s}^{\mathrm{h}}_{mk_l}$ in (\ref{eq:sc_vector_steering}) is ignored when constructing 
$\hat{\mathbf{V}}^{\mathrm{h}}_{k_l}$ in (\ref{eq:sc_horizontal}) if 
\begin{align}
\frac{d^{\mathrm{h}}_{11k_l}}{d^{\mathrm{h}}_{mmk_l}} > \gamma. 
\label{eq:threshold}
\end{align} 
Note that, based on (\ref{eq:sc_vector_steering}) and (\ref{eq:diagonal_steering}), $d^{\mathrm{h}}_{11k_l}$ is the largest one of elements $d^{\mathrm{h}}_{mmk_l}$. Let $N^{\mathrm{h}}_{k_l}$ denote the number of elements $d^{\mathrm{h}}_{mmk_l}$ satisfying  (\ref{eq:threshold}), then only the vectors $\mathbf{s}^{\mathrm{h}}_{mk_l}$ associated with the $N^{\mathrm{h}}_{k_l}$ largest elements $d^{\mathrm{h}}_{mmk_l}$ are employed to constructed $\hat{\mathbf{V}}^{\mathrm{h}}_{k_l}$ in (\ref{eq:sc_horizontal}). In addition, based on (\ref{eq:rank_horizontal}), the rank of $\hat{\mathbf{V}}^{\mathrm{h}}_{k_l}$ in (\ref{eq:sc_horizontal}) should be no less than $\kappa_\mathrm{h}$ in (\ref{eq:rank_horizontal}). Hence, define
\begin{align}
\breve{N}^{\mathrm{h}}_{k_l} = \min\left\lbrace N^{\mathrm{h}}_{k_l}, \kappa_\mathrm{h} \right\rbrace.
\label{eq:number_vector_horizontal}
\end{align} 
Then, (\ref{eq:sc_horizontal}) can be revised as 
\begin{align}
\hat{\mathbf{V}}^{\mathrm{h}}_{k_l} = \frac{1}{ \left\|\breve{\mathbf{D}}_{k_l}^\mathrm{h} \right\|_\mathrm{F}
 }  \breve{\mathbf{S}}^{\mathrm{h}}_{k_l} \breve{\mathbf{D}}_{k_l}^\mathrm{h},
\label{eq:sc_horizontal_reduce}
\end{align} 
where the $\breve{N}^{\mathrm{h}}_{k_l} \times \breve{N}^{\mathrm{h}}_{k_l}$ diagonal matrix $\breve{\mathbf{D}}_{k_l}^\mathrm{h}$ includes only the $\breve{N}^{\mathrm{h}}_{k_l}$ largest elements $d^{\mathrm{h}}_{mmk_l}$, and the $M_\mathrm{h} \times \breve{N}^{\mathrm{h}}_{k_l}$ matrix $\breve{\mathbf{S}}^{\mathrm{h}}_{k_l}$ includes only the vectors $\mathbf{s}^{\mathrm{h}}_{mk_l}$ associated with the $\breve{N}^{\mathrm{h}}_{k_l}$ largest elements $d^{\mathrm{h}}_{mmk_l}$. Similar dimension reduction can be achieved for 
the 
vertical subspace as well.

\subsection{Algorithm Robustness} \label{subsec:robustness}

With channel estimation error, the estimated 
instantaneous CSI $\mathbf{h}_{{k_l}l}^\mathrm{h}$ in (\ref{eq:csi_horizontal}) is modeled as in \cite{Rusek_massive_MIMO_overview} as 
\begin{align}
\hat{\mathbf{h}}_{{k_l}l}^\mathrm{h} = \zeta_{k_l} \mathbf{h}_{{k_l}l}^\mathrm{h} + \sqrt{1-\zeta_{k_l}^2} \boldsymbol{\epsilon}_{{k_l}l},
\label{eq:est_csi_horizontal}
\end{align}
where $\zeta_{k_l} \in [0,1]$ is the CSI accuracy level for the associated UE and the $M_\mathrm{h} \times 1$ vector $\boldsymbol{\epsilon}_{{k_l}l}$ with each element being a zero-mean unit-variance complex-valued Gaussian random variable denotes the error vector. Then, based on (\ref{eq:cov_horizontal}), the $M_\mathrm{h} \times M_\mathrm{h}$ 
SCM based on $\hat{\mathbf{h}}_{{k_l}l}^\mathrm{h}$ for the horizontal dimension $\hat{\mathbf{R}}_{k_l}^\mathrm{h}$ is 
\begin{align}
\hat{\mathbf{R}}_{k_l}^\mathrm{h} 
= \mathrm{E} \left[ \hat{\mathbf{h}}_{{k_l}l}^\mathrm{h} \left( \hat{\mathbf{h}}_{{k_l}l}^\mathrm{h} \right)^\mathrm{H} \right] 
\approx \zeta_{k_l}^2 \mathbf{R}_{k_l}^\mathrm{h} + \left(1-\zeta_{k_l}^2 \right) \mathbf{I}_{M_\mathrm{h}}.
\label{eq:est_cov_horizontal}
\end{align}  
Then, based on (\ref{eq:dominant_search}) and (\ref{eq:est_cov_horizontal}), the estimated dominant angle $\hat{\omega}_{0k_l}^\mathrm{h} \in [0,2\pi)$ is
\ifCLASSOPTIONonecolumn  
\begin{align}
\hat{\omega}_{0k_l}^\mathrm{h} 
=  \mathop {\arg\max} \limits_{\omega \in \left[0,2\pi \right)} \mathbf{e}_{M_\mathrm{h}}^\mathrm{H} \left[\omega \right] \hat{\mathbf{R}}_{k_l}^\mathrm{h} \mathbf{e}_{M_\mathrm{h}}\left[\omega \right] 
\approx 
\mathop {\arg\max} \limits_{\omega \in \left[0,2\pi \right)} \left\lbrace \zeta_{k_l}^2 \mathbf{e}_{M_\mathrm{h}}^\mathrm{H} \left[\omega \right] \mathbf{R}_{k_l}^\mathrm{h}  \mathbf{e}_{M_\mathrm{h}}\left[\omega \right] + 1-\zeta_{k_l}^2 \right\rbrace 
.
\label{eq:est_dominant_search}
\end{align}  
\else
\begin{align}
\hat{\omega}_{0k_l}^\mathrm{h} & =  \mathop {\arg\max} \limits_{\omega \in \left[0,2\pi \right)} \mathbf{e}_{M_\mathrm{h}}^\mathrm{H} \left[\omega \right] \hat{\mathbf{R}}_{k_l}^\mathrm{h} \mathbf{e}_{M_\mathrm{h}}\left[\omega \right] \nonumber \\ 
& \approx 
\mathop {\arg\max} \limits_{\omega \in \left[0,2\pi \right)} \left\lbrace \zeta_{k_l}^2 \mathbf{e}_{M_\mathrm{h}}^\mathrm{H} \left[\omega \right] \mathbf{R}_{k_l}^\mathrm{h}  \mathbf{e}_{M_\mathrm{h}}\left[\omega \right] + 1-\zeta_{k_l}^2 \right\rbrace 
.
\label{eq:est_dominant_search}
\end{align} 
\fi
When $\zeta_{k_l} = 0$, (\ref{eq:est_dominant_search}) is unsolvable. Otherwise, (\ref{eq:est_dominant_search}) becomes
\begin{align}
\hat{\omega}_{0k_l}^\mathrm{h} \approx \omega_{0k_l}^\mathrm{h}.
\label{eq:est_dominant_search_robust}
\end{align}  
Note that in practice, (\ref{eq:est_dominant_search_robust}) is generally hold, i.e., the derived dominant angle $\hat{\omega}_{0k_l}^\mathrm{h}$ to construct the subspace is almost the same as $\omega_{0k_l}^\mathrm{h}$ in (\ref{eq:dominant_search}) derived with perfect CSI, unless $\zeta_{k_l} \rightarrow 0$ or the number of used subcarriers for the expectation in (\ref{eq:est_cov_horizontal}) is too small. In other words, the proposed SC method is very robust against inaccurate CSI. This robustness is also valid for the vertical dimension. 

\subsection{Complexity} \label{subsec:complexity_sc}
The complexity levels of (\ref{eq:cov_horizontal}), (\ref{eq:dominant_search}), (\ref{eq:diagonal_steering}), and (\ref{eq:sc_horizontal_reduce}) are $\mathcal{O}(M_{\mathrm{h}}^2)$, $\mathcal{O}[N_{\mathrm{F}}^{\mathrm{h}}\log_2({N_{\mathrm{F}}^{\mathrm{h}}})]$, $\mathcal{O}(M_{\mathrm{h}}^2)$, and $\mathcal{O}(M_{\mathrm{h}}\breve{N}^{\mathrm{h}}_{k_l})$ respectively for the horizontal dimension with the dimension reduction method presented in Section \ref{subsec:dimension}. For (\ref{eq:sc_total}) and (\ref{eq:cov_est}), the complexity levels are $\mathcal{O}(M\breve{N}^{\mathrm{h}}_{k_l}\breve{N}^{\mathrm{v}}_{k_l})$ and $\mathcal{O}(M^2\breve{N}^{\mathrm{h}}_{k_l}\breve{N}^{\mathrm{v}}_{k_l})$ respectively, where $\breve{N}^{\mathrm{v}}_{k_l}$ is the counterpart of $\breve{N}^{\mathrm{h}}_{k_l}$ defined in (\ref{eq:number_vector_horizontal}) for the vertical dimension. Because $M$ is large in massive MIMO systems, the complexity of $\mathcal{O}(M^2\breve{N}^{\mathrm{h}}_{k_l}\breve{N}^{\mathrm{v}}_{k_l})$ for (\ref{eq:cov_est}) is the dominant complexity. Therefore, the complexity to derive $\mathbf{R}$ in (\ref{eq:eigen}) for all $K$ UEs is $\mathcal{O}(M^2K)$. Since the change of spatial subspace is relatively slow, the estimated $\mathbf{R}$ can be used for the HB method proposed in Section \ref{sec:hybrid} for a relatively long time before being updated. As a result, the complexity is in fact very low.  

\section{Subspace Construction in FDD Systems} \label{sec:fdd}
In Section \ref{sec:subspace}, a novel SC method based on partial uplink instantaneous CSI is proposed to estimate the SCM $\mathbf{R}_{k_l}$ for the proposed HB in Section {\ref{sec:hybrid} in TDD systems for both the uplink and downlink, exploiting the channel reciprocity. Since it is based on partial uplink CSI, it can also be applied for the uplink of FDD systems. However, it cannot be directly applied to the downlink of FDD systems because the uplink and downlink channels are not reciprocal. For conventional FDD systems, the downlink CSI is normally measured at each UE by the downlink pilots transmitted by the BS and the measured CSI of each UE is then fed back to the BS \cite{Adhikary_JSDM,Jiang_FDD_Massive_MIMO}. In the case of massive MIMO systems, this downlink channel training and feedback scheme may cause excessive channel estimation overhead. 
With HB, since $R \ll M$, the conventional channel training and feedback scheme could be still employed to estimate the effective frequency-domain downlink channel matrix at the baseband as $(\mathbf{H}^{\mathrm{BB}}_l)^\mathrm{T}$ in (\ref{eq:io_bb}) once the first-level unified AB matrix $\mathbf{W}^{\mathrm{RF}}$ in (\ref{eq:precoding}) is determined, and to construct the baseband digital precoding matrix as $\mathbf{W}_l^{\mathrm{BB}}$ in (\ref{eq:precoding_bb}). However, 
to apply the unified AB proposed in Section \ref{subsec:analog} for the downlink of FDD systems, the BS needs to estimate the downlink SCM $\mathbf{R}^{\mathrm{DL}}_{k_l}$. 
In this section, the application of the SC method proposed in Section \ref{sec:subspace} 
to the downlink of FDD massive MIMO-OFDM systems is briefly discussed. 

\subsection{Based on Partial Downlink CSI} \label{subsec:fdd_dcsi}
One straightforward way to apply the proposed SC method based on partial downlink CSI for the downlink of FDD systems is described as below. 
\begin{enumerate}
\item{The BS transmits downlink pilots such as the cyclic shift method based on ZC sequences employed in LTE \cite{3GPP_TS_36.211} from only $P = M_\mathrm{h}+M_\mathrm{v}-1$ antennas as in Section \ref{subsec:subspace} to UEs.}
\item{Each UE acquires the required partial downlink CSI defined in Section \ref{subsec:subspace}
.}
\item{Each UE feeds back the required partial downlink CSI to the BS.}
\item{The BS applies the proposed SC method to estimate the downlink SCM 
$\mathbf{R}^{\mathrm{DL}}_{k_l}$
.}
\end{enumerate}
For this method, each UE needs to feed back a $P\times 1$ partial instantaneous CSI vector to the BS, then the whole proposed SC algorithm is carried out at the BS. 

An alternative method is to replace the third and forth steps of the above method as described below.
\begin{enumerate}
\setcounter{enumi}{2}
\item{Each UE applies the proposed SC method until (\ref{eq:threshold}), then it feeds back the $\breve{N}^\mathrm{h}_{k_l}$ largest elements $d^\mathrm{h}_{mmk_l}$ in (\ref{eq:diagonal_steering}) where $\breve{N}^\mathrm{h}_{k_l}$ is defined in (\ref{eq:number_vector_horizontal}), their associated indices, and 
$\omega^\mathrm{h}_{0k_l}$ in (\ref{eq:dominant_search_fft}), to the BS for the horizontal dimension. Similar parameters are also acquired at each UE and fed back to the BS for the vertical dimension
.}
\item{The BS applies (\ref{eq:sc_horizontal_reduce}) for the horizontal dimension and the corresponding equation for the vertical dimension
, then $\mathbf{R}^{\mathrm{DL}}_{k_l}$ is estimated as (\ref{eq:sc_total}).} 
\end{enumerate}
For the alternative method, each UE needs to feed back $\breve{N}^\mathrm{h}_{k_l}$ largest elements $d^\mathrm{h}_{mmk_l}$, their $(\breve{N}^\mathrm{h}_{k_l}-1)$ associated indices since $d^\mathrm{h}_{11k_l}$ is the largest one for certain, and the dominant angle 
$\omega^\mathrm{h}_{0k_l}$, for the horizontal dimension, as well as the associated  parameters for the vertical dimension. Compared to the first method, this method may have lower overhead, but part of the proposed SC algorithm needs to be carried out at each UE. 
Note that the two proposed methods for the downlink of FDD systems in this subsection can also be applied to the downlink of TDD systems.

\subsection{Based on Partial Uplink CSI} \label{subsec:fdd_ucsi}
For FDD systems, although the downlink and uplink channels are not reciprocal, under the FDD channel reciprocity model in \cite{R1-151920,3GPP_TR_36.897}, they still share same spatial features, i.e., in (\ref{eq:channel_time}), $|\tilde{\alpha}_{ijk}^\mathrm{DL}| = |\tilde{\alpha}_{ijk}^\mathrm{UL}|$, $\phi^\mathrm{DL}_{ijk}=\phi^\mathrm{UL}_{ijk}$, $\theta^\mathrm{DL}_{ijk} = \theta^\mathrm{UL}_{ijk}$, and $\tau^\mathrm{DL}_{ijk} = \tau^\mathrm{UL}_{ijk}$
. In other words, the non-reciprocity is mainly caused by different 
$\mathbf{a}^\mathrm{DL}(\phi,\theta)$ and $\mathbf{a}^\mathrm{UL}(\phi,\theta)$ 
in (\ref{eq:channel_time})
, due to different wavelengths of the downlink and uplink carrier frequencies according to (\ref{eq:upa})-(\ref{eq:upa_vertical}). Note that (\ref{eq:upa_horizontal}) and (\ref{eq:upa_vertical}) can be represented respectively as 
\begin{align}
\mathbf{a}_{\mathrm{h}}\left(\phi,\theta\right) = \mathbf{e}_{M_\mathrm{h}} \left[\omega_{\mathrm{h}} = 2\pi\frac{d_\mathrm{h}}{\lambda}\cos{\phi}\sin{\theta} \right] ,
\label{eq:upa_horizontal_angle}
\end{align}
and
\begin{align}
\mathbf{a}_{\mathrm{v}}\left(\theta\right) = \mathbf{e}_{M_\mathrm{v}} \left[\omega_{\mathrm{v}} = 2\pi\frac{d_\mathrm{v}}{\lambda}\cos{\theta} \right],
\label{eq:upa_vertical_angle}
\end{align}
where $\mathbf{e}_M [\omega]$ is defined in (\ref{eq:vector_steering}). Substituting the downlink wavelength $\lambda^{\mathrm{DL}}$ and the uplink wavelength $\lambda^{\mathrm{UL}}$ into $\omega_\mathrm{h}$ and $\omega_\mathrm{v}$ in (\ref{eq:upa_horizontal_angle}) and (\ref{eq:upa_vertical_angle}) respectively, the  relations between the corresponding $\omega_\mathrm{h}^\mathrm{DL}$, $\omega_\mathrm{v}^\mathrm{DL}$, and $\omega_\mathrm{h}^\mathrm{UL}$, $\omega_\mathrm{v}^\mathrm{UL}$, can be written respectively as
\begin{align}
\omega_{\mathrm{h}}^\mathrm{DL} = 2\pi\frac{d_\mathrm{h}}{\lambda^\mathrm{DL}}\cos{\phi}\sin{\theta} = \frac{\lambda^\mathrm{UL}}{\lambda^\mathrm{DL}}\omega_{\mathrm{h}}^\mathrm{UL} = \frac{F_\mathrm{c}^\mathrm{DL}}{F_\mathrm{c}^\mathrm{UL}}\omega_{\mathrm{h}}^\mathrm{UL},
\label{eq:dl_ul_horizontal}
\end{align}
and
\begin{align}
\omega_{\mathrm{v}}^\mathrm{DL} = 2\pi\frac{d_\mathrm{v}}{\lambda^\mathrm{DL}}\cos{\theta} = \frac{\lambda^\mathrm{UL}}{\lambda^\mathrm{DL}}\omega_{\mathrm{v}}^\mathrm{UL} = \frac{F_\mathrm{c}^\mathrm{DL}}{F_\mathrm{c}^\mathrm{UL}}\omega_{\mathrm{v}}^\mathrm{UL},
\label{eq:dl_ul_vertical}
\end{align} 
where $F_\mathrm{c}^\mathrm{DL}$ and $F_\mathrm{c}^\mathrm{UL}$ are the carrier frequencies for the downlink and uplink respectively.

As discussed above, the proposed SC method can be applied for the downlink of FDD systems based on partial uplink CSI as described below. 
\begin{enumerate}
\item{Each UE transmits uplink pilots such as the cyclic shift method based on ZC sequences employed in LTE \cite{3GPP_TS_36.211} to the BS.}
\item{The BS acquires the required partial uplink instantaneous CSI from only $P = M_\mathrm{h}+M_\mathrm{v}-1$ antennas defined in Section \ref{subsec:subspace}
.}
\item{The BS applies the proposed SC method until (\ref{eq:threshold}), then the constructed uplink steering matrix $\breve{\mathbf{S}}^\mathrm{h,UL}_{k_l}$ in (\ref{eq:sc_horizontal_reduce}) defined in (\ref{eq:sc_vector_steering}) is reconstructed as $\breve{\mathbf{S}}^\mathrm{h,DL}_{k_l}$ for the downlink with modified downlink steering angles based on (\ref{eq:dl_ul_horizontal}). Similar modifications are also applied to the vertical dimension.}
\item{The BS applies (\ref{eq:sc_horizontal_reduce}) for the horizontal dimension and the corresponding equation for the vertical dimension
, then $\mathbf{R}^{\mathrm{DL}}_{k_l}$ is estimated as (\ref{eq:sc_total}).} 
\end{enumerate}

\section{Simulation Results} \label{sec:results}


In this section, simulation results are provided to investigate the performance of the proposed HB with unified AB and the performance of the proposed SC algorithm with partial CSI. 
Specifically, the TDD carrier frequency is $F_{\mathrm{c}}=2.6\mathrm{GHz}$. As for FDD, the carrier frequency for the downlink is considered as $F_{\mathrm{c}}^\mathrm{DL}=2.6\mathrm{GHz}$, and the carrier frequency for the uplink is considered as $\eta F_{\mathrm{c}}^\mathrm{DL}$, where $\eta = F_{\mathrm{c}}^\mathrm{UL}/F_{\mathrm{c}}^\mathrm{DL}$ is a configurable parameter. The bandwidth $B=20\mathrm{MHz}$ is a particular number in LTE \cite{3GPP_TS_36.104}. The sampling frequency $F_{\mathrm{s}}=30.72\mathrm{MHz}$, the system FFT size $L=2048$, and the number of used subcarriers $L_{\mathrm{used}}=1200$, are the values associated with $B=20\mathrm{MHz}$ in LTE \cite{3GPP_TS_36.104}, and normal CP defined in \cite{3GPP_TS_36.211} is employed. The channel model is considered as the Urban Microcell (UMi) model in \cite{3GPP_TR_36.873}. 
The UE LoS probability and NLoS probability are assumed to be $P_{\mathrm{L}}=50\%$ and $P_{\mathrm{N}}=50\%$ respectively. The UE velocity is $v_{\mathrm{ue}}=3\mathrm{km/h}$. Each channel realization duration is $T_{\mathrm{cr}}=10\mathrm{ms}$, i.e., $140$ OFDM symbols with normal CP in LTE \cite{3GPP_TS_36.211}, which means that the second-order SCM $\mathbf{R}_{k_l}$ is updated every $140$ OFDM symbols. The number of BS antennas is $M=256$, and this $M_\mathrm{h}\times M_\mathrm{v}$ antenna array is placed as $16\times 16$ for UPA and $256\times 1$ for ULA with half wavelength antenna spacing for both dimensions. 
The horizontal angle spread of BS antennas is $\phi\in[\pi/6,5\pi/6]$, which is considered to cover UEs in a sector. As for the vertical angle spread, two cases are simulated, where the narrower spread $\theta\in[5\pi/9,3\pi/4]$ is considered to only cover ground UEs, while the broader spread $\theta\in[\pi/3,3\pi/4]$ is considered to cover the ground UEs as well as UEs on different floors of a tall building. The horizontal and vertical angles of the direct path of a UE is uniformly distributed in the considered horizontal angle spread and vertical angle spread of BS antennas respectively.
The number of UE groups is $N_\mathrm{g}=4$, and the number of UEs per group is $K_l=16$. The effective CSI seen at the baseband, e.g., $(\mathbf{H}_l^{\mathrm{BB}})^{\mathrm{T}}$ in (\ref{eq:io_bb}) for TDD systems, and used for digital beamforming is assumed to be perfect to focus on the performance of proposed algorithms, and the digital beamforming granularity is $L_\mathrm{pg}=12$, which means that only one digital beamforming matrix is calculated and employed for $L_\mathrm{pg}=12$ continuous subcarriers, in order to shorten the simulation time.

\subsection{Proposed HB} \label{subsec:results_hb}
\ifCLASSOPTIONonecolumn
\begin{figure}[!t]
\centering \includegraphics[width = 1.0\linewidth]{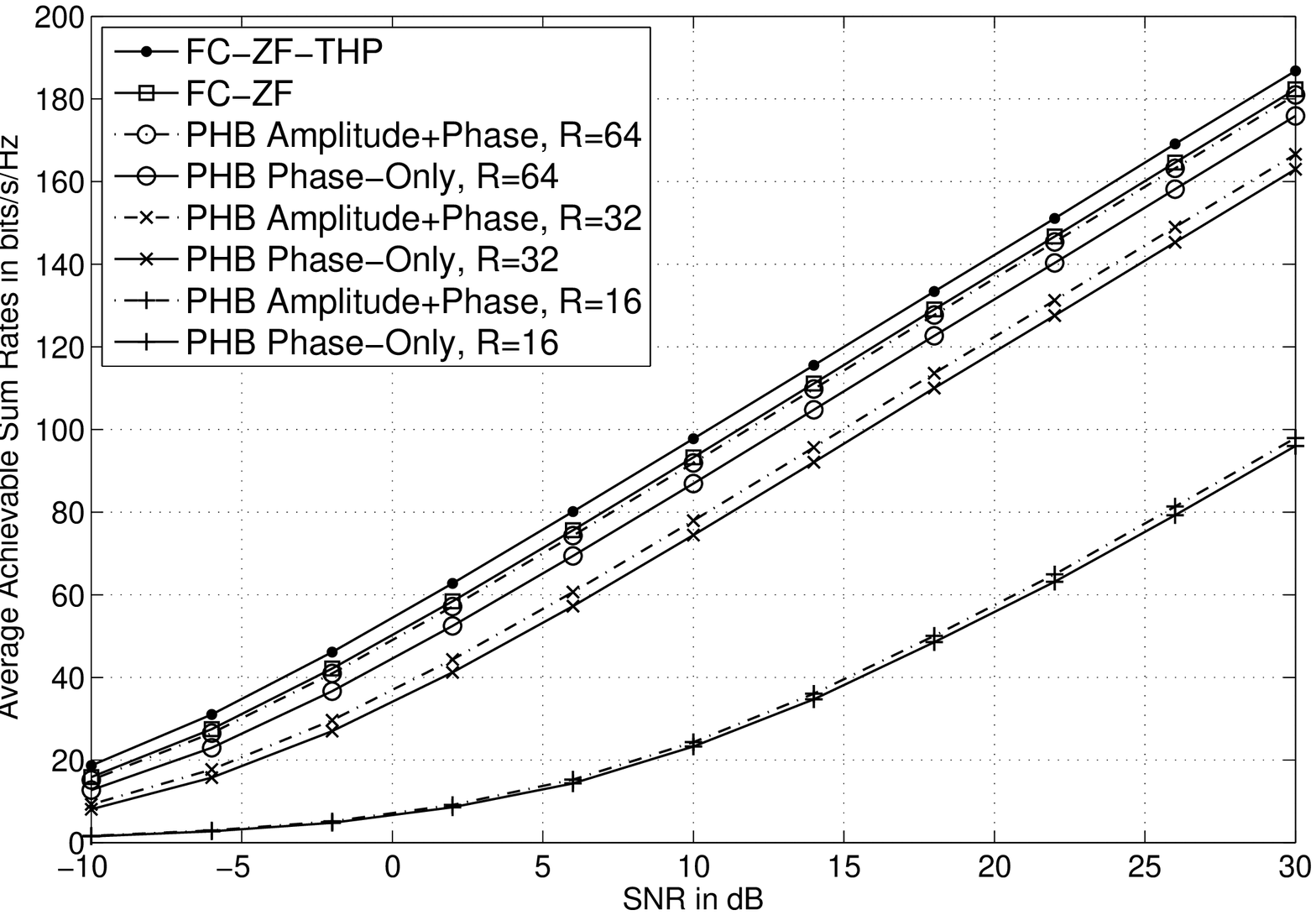}
\caption{Downlink average achievable sum rates of the proposed HB method with perfect SCM knowledge for a $16\times 16$ UPA with $\phi\in[\pi/6,5\pi/6]$ and $\theta\in[5\pi/9,3\pi/4]$, employing different $R$ values under different SNRs in a TDD system.} 
\label{fig:results_upa_hb35}
\end{figure}

\begin{figure}[!t]
\centering \includegraphics[width = 1.0\linewidth]{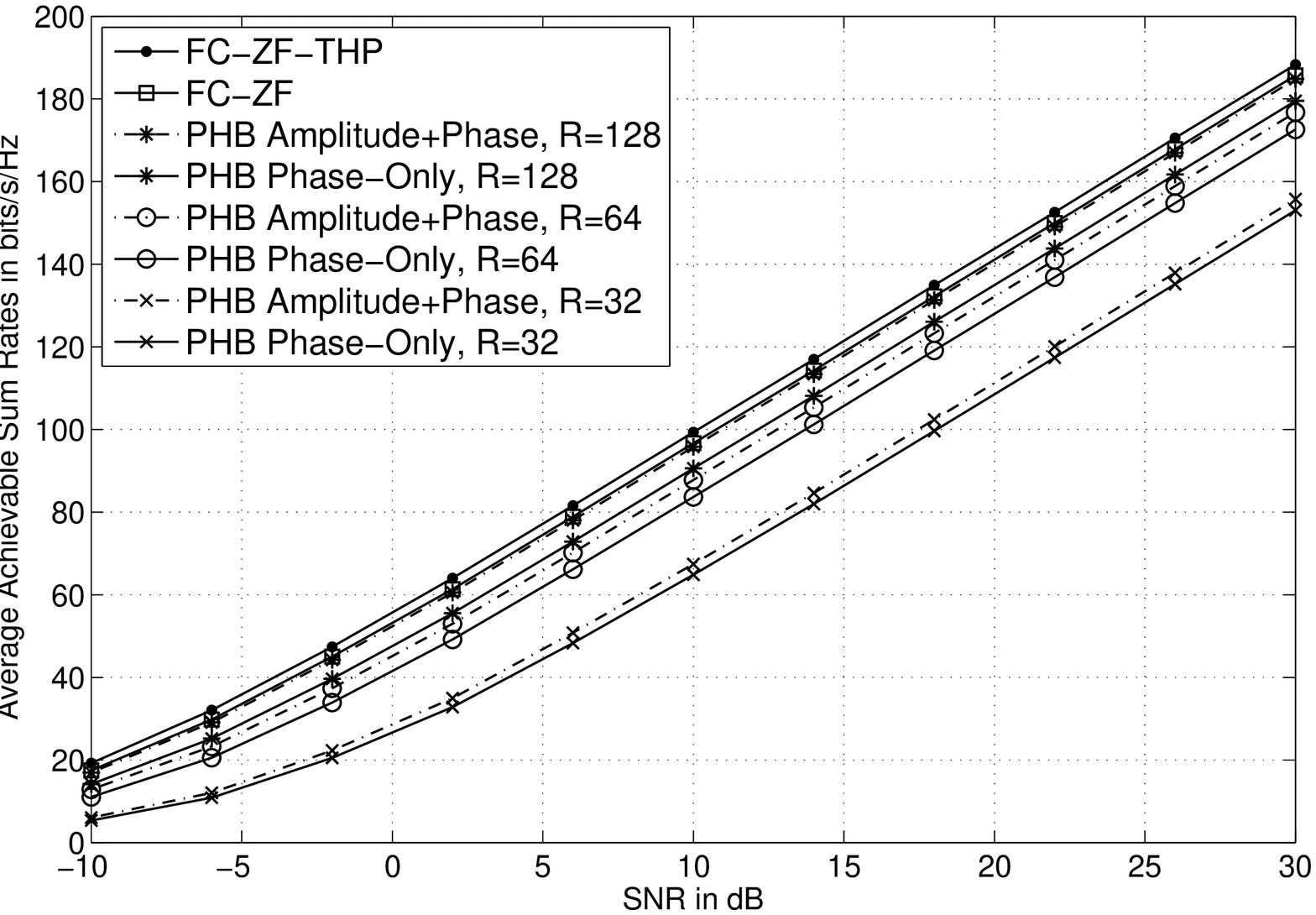}
\caption{Downlink average achievable sum rates of the proposed HB method with perfect SCM knowledge for a $16\times 16$ UPA with $\phi\in[\pi/6,5\pi/6]$ and $\theta\in[\pi/3,3\pi/4]$, employing different $R$ values under different SNRs in a TDD system.}
\label{fig:results_upa_hb75}
\end{figure}

\begin{figure}[!t]
\centering \includegraphics[width = 1.0\linewidth]{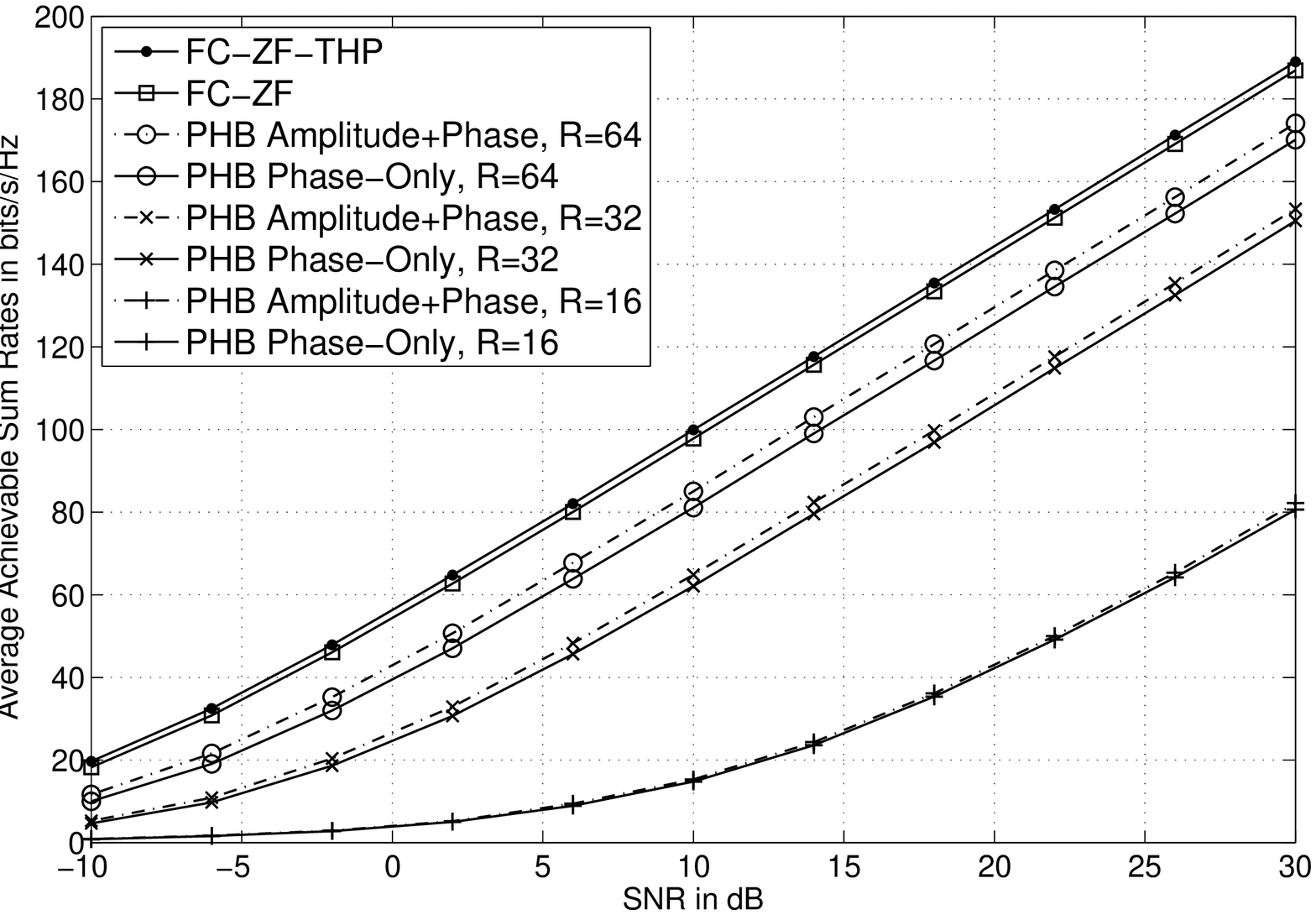}
\caption{Downlink average achievable sum rates of the proposed HB method with perfect SCM knowledge for a $256\times 1$ ULA with $\phi\in[\pi/6,5\pi/6]$ and $\theta\in[5\pi/9,3\pi/4]$, employing different $R$ values under different SNRs in a TDD system.}
\label{fig:results_ula_hb35}
\end{figure}
\else
\begin{figure}[!t]
\centering \includegraphics[width = 0.95\linewidth]{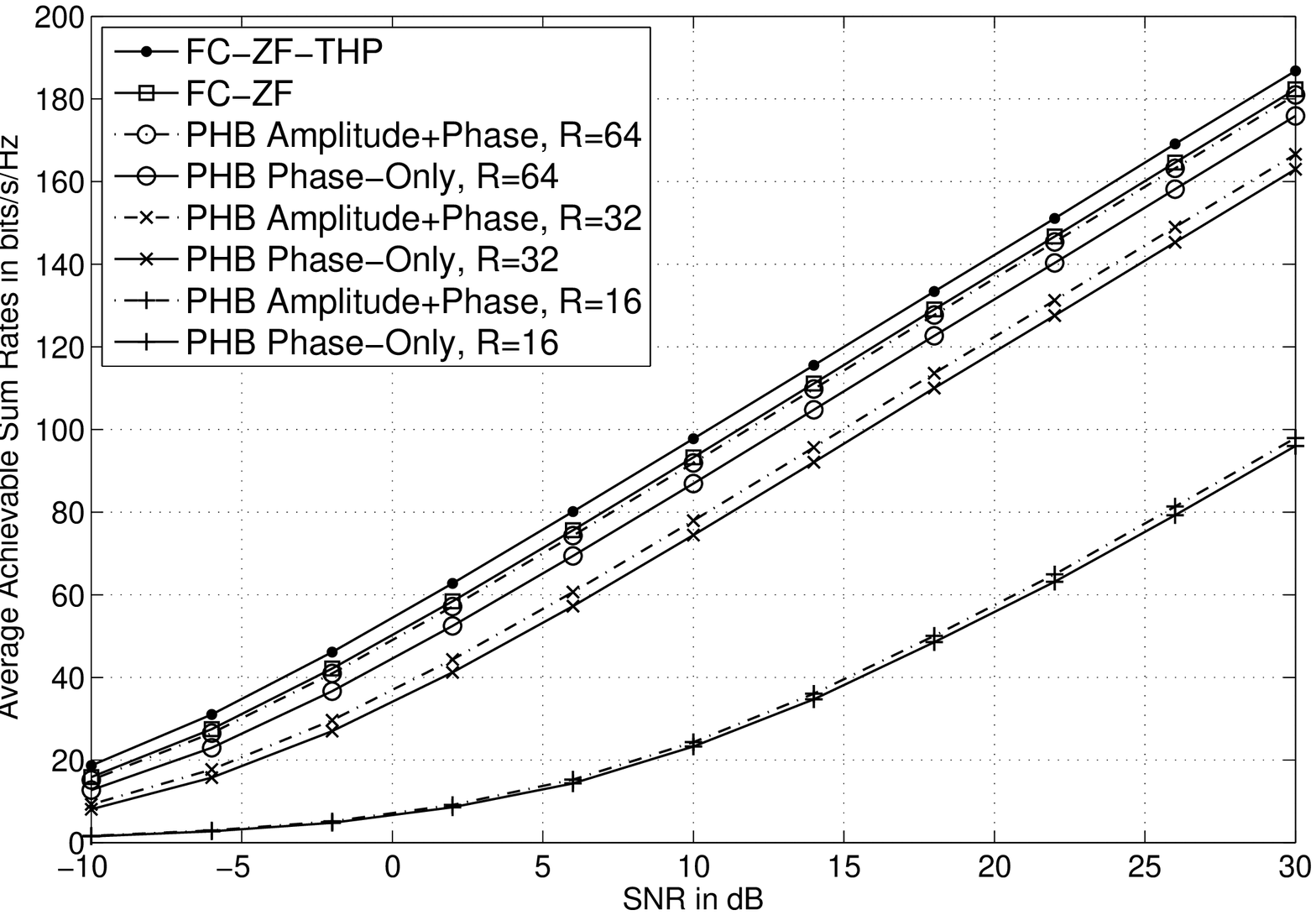}
\caption{Downlink average achievable sum rates of the proposed HB method with perfect SCM knowledge for a $16\times 16$ UPA with $\phi\in[\pi/6,5\pi/6]$ and $\theta\in[5\pi/9,3\pi/4]$, employing different $R$ values under different SNRs in a TDD system.} 
\label{fig:results_upa_hb35}
\end{figure}

\begin{figure}[!t]
\centering \includegraphics[width = 0.95\linewidth]{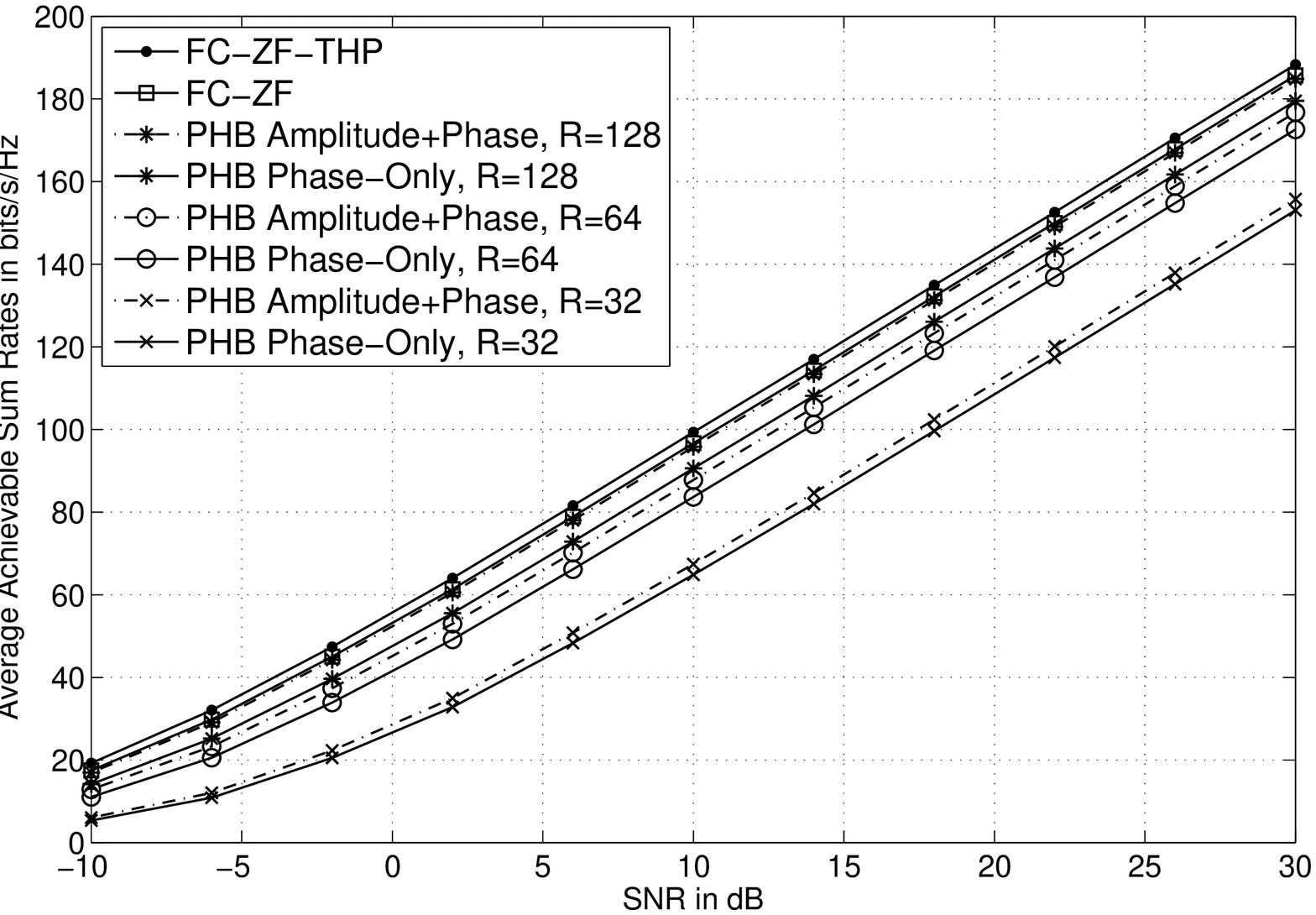}
\caption{Downlink average achievable sum rates of the proposed HB method with perfect SCM knowledge for a $16\times 16$ UPA with $\phi\in[\pi/6,5\pi/6]$ and $\theta\in[\pi/3,3\pi/4]$, employing different $R$ values under different SNRs in a TDD system.}
\label{fig:results_upa_hb75}
\end{figure}

\begin{figure}[!t]
\centering \includegraphics[width = 0.95\linewidth]{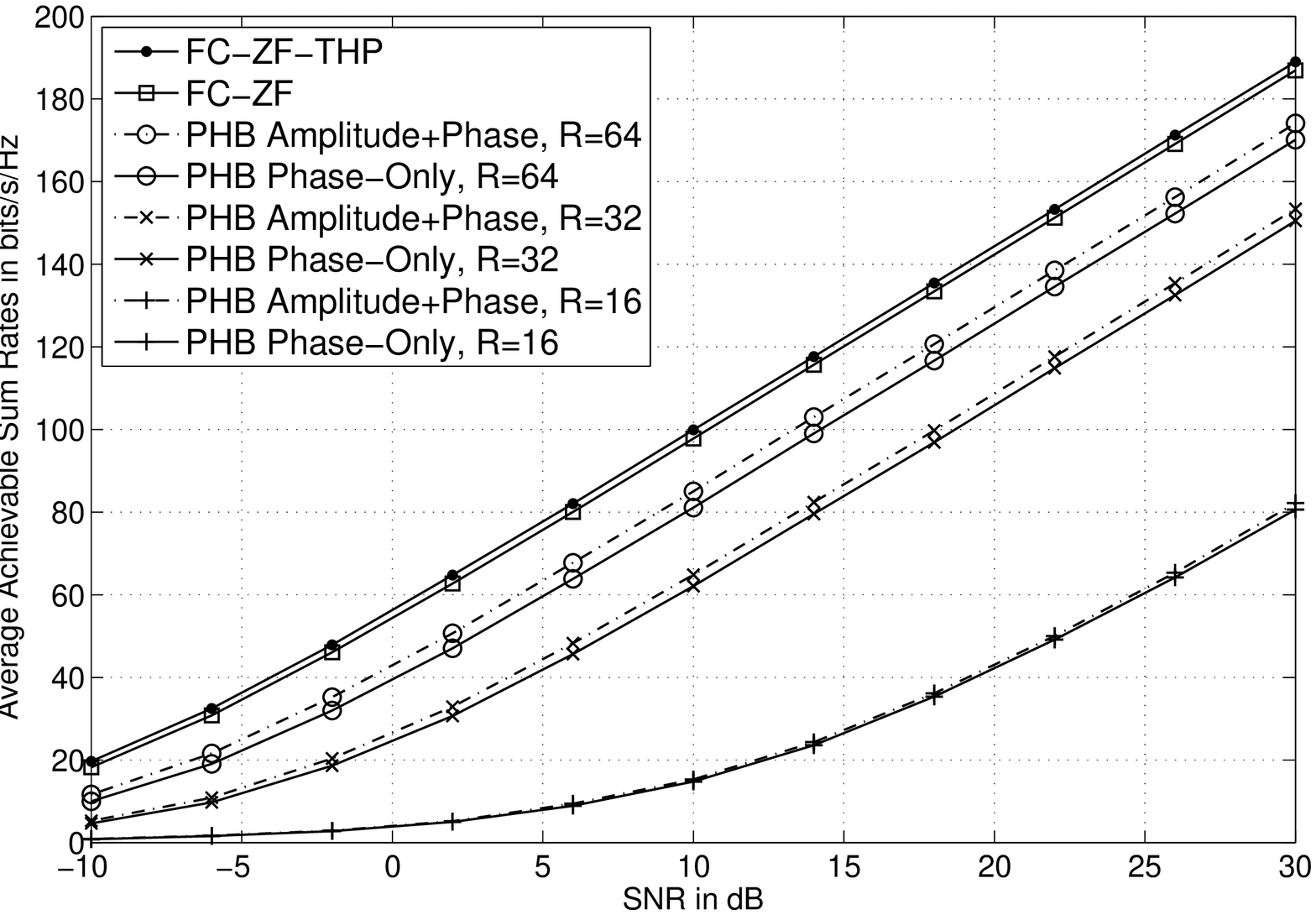}
\caption{Downlink average achievable sum rates of the proposed HB method with perfect SCM knowledge for a $256\times 1$ ULA with $\phi\in[\pi/6,5\pi/6]$ and $\theta\in[5\pi/9,3\pi/4]$, employing different $R$ values under different SNRs in a TDD system.}
\label{fig:results_ula_hb35}
\end{figure}
\fi
In this subsection, the performance of the proposed HB with unified AB presented in Section \ref{sec:hybrid} is investigated for TDD massive MIMO-OFDM systems assuming that the BS knows the second-order SCM knowledge of all UEs. Three cases are considered, i.e., a $16\times 16$ UPA with $\phi\in[\pi/6,5\pi/6]$ and $\theta\in[5\pi/9,3\pi/4]$, a $16\times 16$ UPA with $\phi\in[\pi/6,5\pi/6]$ and $\theta\in[\pi/3,3\pi/4]$, and a $256\times 1$ ULA with $\phi\in[\pi/6,5\pi/6]$ and $\theta\in[5\pi/9,3\pi/4]$, and their downlink average achievable sum rates are presented in Fig. \ref{fig:results_upa_hb35} to Fig. \ref{fig:results_ula_hb35} respectively with different $R$ values under different SNRs per subcarrier. 
In the figures, the results for the fully digital capacity-approaching ZF-THP and the fully digital ZF based on full CSI with $M=256$ RF chains are shown as references, which are denoted as FC-ZF-THP and FC-ZF in the figures respectively. Based on Section \ref{subsec:rf_chains}, the analyzed sufficient numbers of $R$ in order to achieve the performance close to FC-ZF are $\kappa_\mathrm{h} \times \kappa_\mathrm{v} \approx 14 \times 5 = 70 \approx 64$, $\kappa_\mathrm{h} \times \kappa_\mathrm{v} \approx 14 \times 10 = 140 \approx 128$, and $\kappa_\mathrm{h} \times \kappa_\mathrm{v} \approx 224 \times 1 = 224$ for the three considered cases respectively. 
For the proposed HB, other than the results with the proposed unified AB realized by a PSN that can only adjust phases in (\ref{eq:solution_relax}), denoted as PHB Phase-Only in the figures, the results with the ideal unified AB that can adjust both amplitudes and phases in (\ref{eq:solution}), denoted as PHB Amplitude+Phase in the figures, are also presented. Since perfect SCM knowledge of all UEs and perfect effective CSI seen at the baseband for digital beamforming are assumed in simulations, 
the results in Fig. \ref{fig:results_upa_hb35} to Fig. \ref{fig:results_ula_hb35} are also valid for the downlink of FDD systems with the same carrier frequency. Note that only the results for the proposed HB are presented in this subsection because prior HB methods cannot support multiple group of UEs which is our considered case as explained in Section \ref{sec:introduction}. More explanation is provided in Section \ref{subsec:discussion}.

Based on the results in Fig. \ref{fig:results_upa_hb35} to Fig. \ref{fig:results_ula_hb35}, several observations are listed below.
\begin{enumerate}
\item{For all three cases, FC-ZF can achieve more than $97{\%}$
performance of the capacity approaching FC-ZF-THP, which is consistent with \cite{Rusek_massive_MIMO_overview,Ngo_Favorable}.}
\item{Fig. \ref{fig:results_upa_hb35} and Fig. \ref{fig:results_upa_hb75} verify 
that with the sufficient number of $R$ analyzed in Section \ref{subsec:rf_chains}, PHB Amplitude+Phase can achieve almost the same performance as FC-ZF. Otherwise, the performance suffers degradation, and a smaller number of $R$ results in more degradation. In addition, with a narrower angle spread, i.e., higher antenna directivity, a smaller $R$ is required to achieve almost the same performance as FC-ZF.}
\item{For all three cases, PHB Phase-only can achieve more
than $95\%$ performance of PHB Amplitude+Phase, and the 
 difference becomes smaller as 
 $R$ reduces. The results indicate that although adjusting both amplitudes and phases for the unified AB can achieve better performance compared to the phase-only PSN, the 
 gain is very limited. In other words, the PSN that can offer
more than $95\%$ performance of FC-ZF is good enough hence well suited for the proposed HB algorithm.}
\item{Fig. \ref{fig:results_upa_hb35} and Fig. \ref{fig:results_ula_hb35} show that with the same $R$ and $M$ values and the same ranges of spread angles, the performance loss from the proposed HB to FC-ZF for ULA is much larger than UPA, which verifies the analysis in Section \ref{subsec:rf_chains} that UPA is better suited for the proposed HB than ULA.}
\item{For all three cases, although performance loss is caused when the number of $R$ is smaller than the sufficient number analyzed in Section \ref{subsec:rf_chains}, acceptable performance could still be achieved as long as $R$ is not too small. As a result, in practice, it is an alternative choice to trade off small performance loss with lower system complexity.}
\end{enumerate}   

\subsection{Proposed SC Method} \label{subsec:results_sc}
\ifCLASSOPTIONonecolumn
\begin{figure}[!t]
\centering \includegraphics[width = 1.0\linewidth]{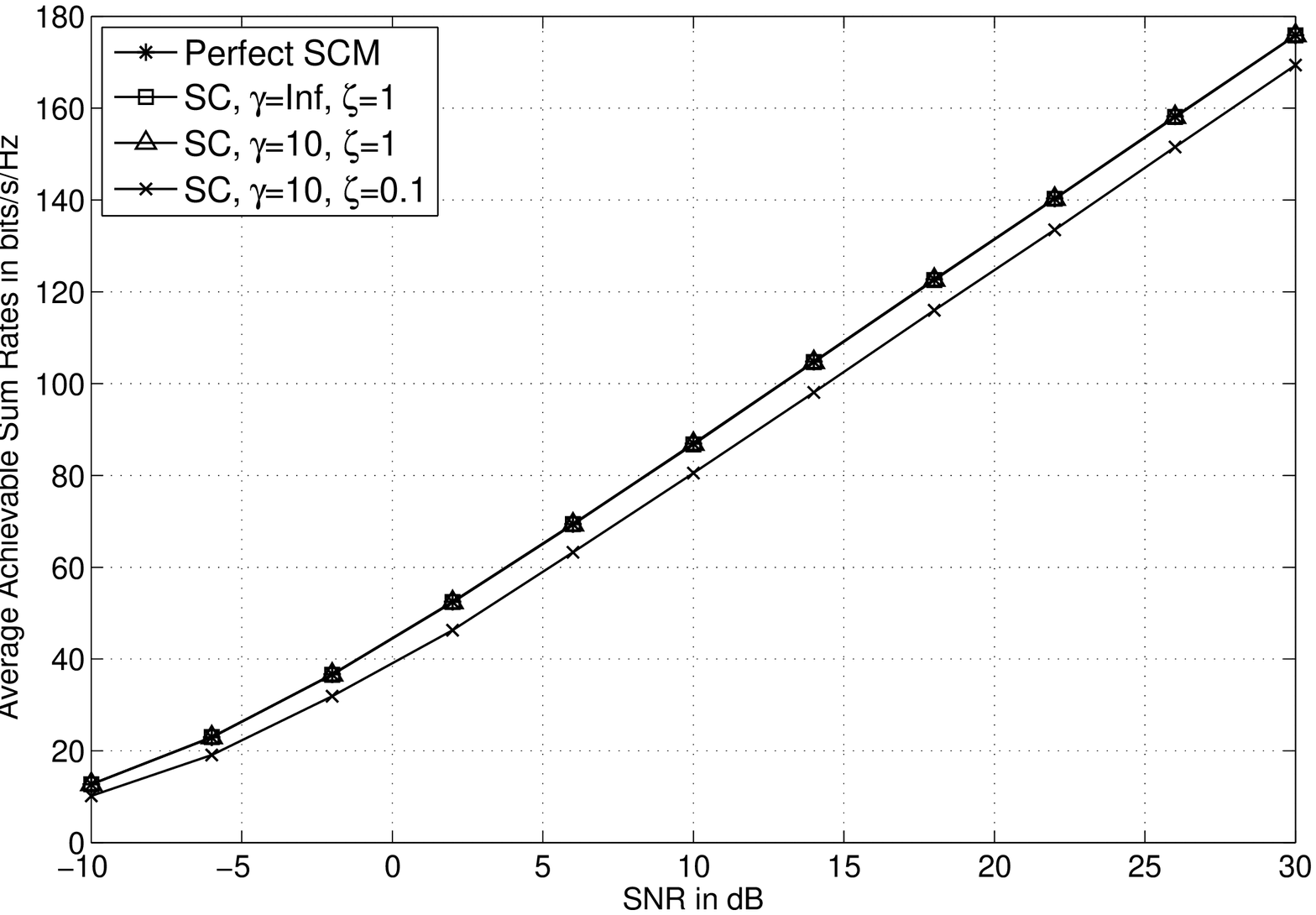}
\caption{Downlink average achievable sum rates of the proposed HB method based on the proposed SC method for a $16\times 16$ UPA with $\phi\in[\pi/6,5\pi/6]$, $\theta\in[5\pi/9,3\pi/4]$, and $R=64$, employing different $\gamma$ and $\zeta$ values under different SNRs in a TDD system.} 
\label{fig:results_upa_sc35}
\end{figure}
\else
\begin{figure}[!t]
\centering \includegraphics[width = 0.95\linewidth]{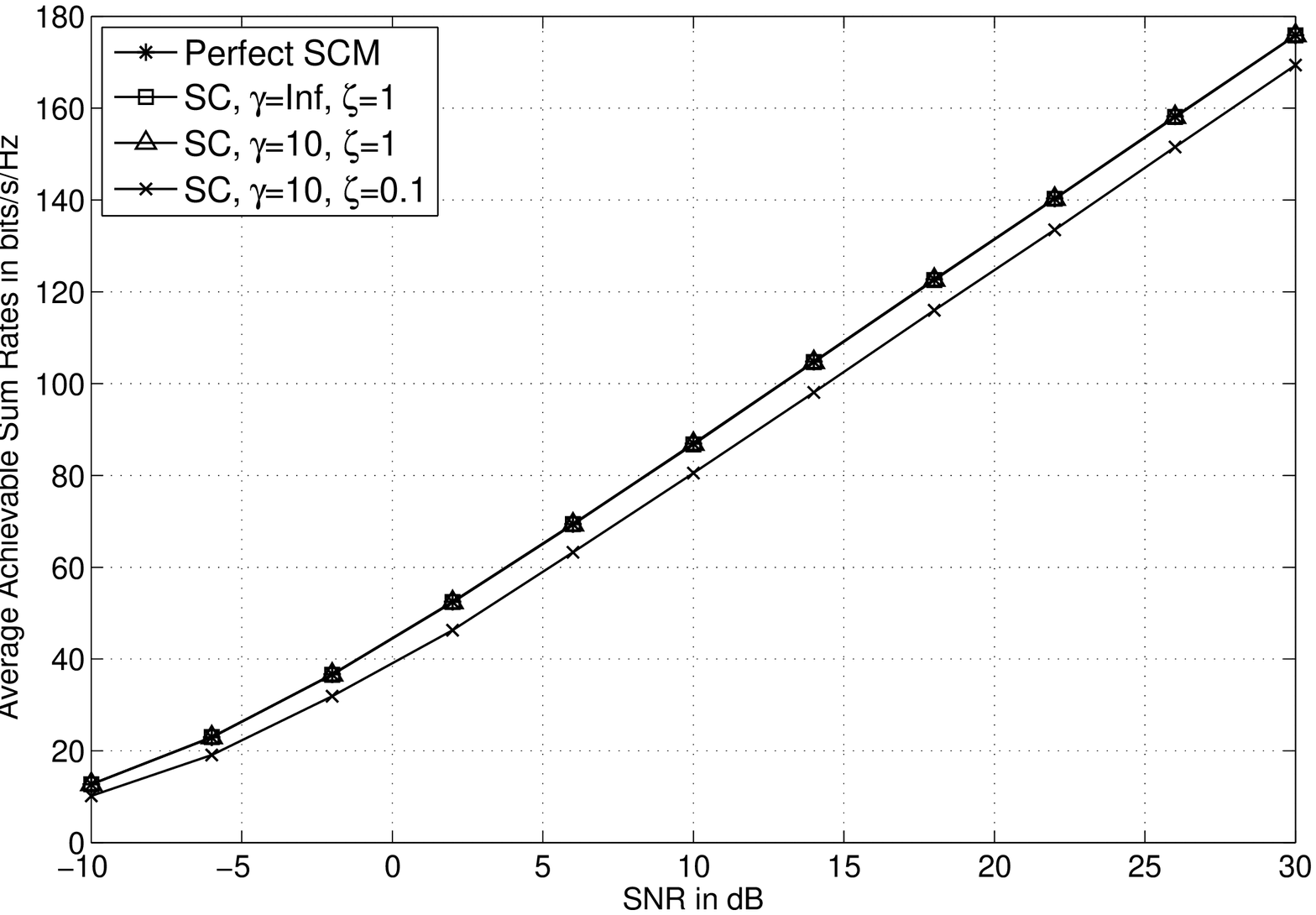}
\caption{Downlink average achievable sum rates of the proposed HB method based on the proposed SC method for a $16\times 16$ UPA with $\phi\in[\pi/6,5\pi/6]$, $\theta\in[5\pi/9,3\pi/4]$, and $R=64$, employing different $\gamma$ and $\zeta$ values under different SNRs in a TDD system.} 
\label{fig:results_upa_sc35}
\end{figure}
\fi


In this subsection, the performance of the proposed SC method 
based on partial CSI presented in Section \ref{sec:subspace} is investigated. In simulations, the threshold $\gamma$ defined in (\ref{eq:threshold}) is selected as $\gamma = \infty$ and $\gamma = 10$. Every $1$ out of $48$ subcarriers, hence $25$ subcarriers in total, are used for the expectation in (\ref{eq:cov_horizontal}) or (\ref{eq:est_cov_horizontal}). For the sake of simplicity, the channel estimation accuracy level $\zeta_{k_l}$ in (\ref{eq:est_csi_horizontal}) is assumed to be the same for all UEs, i.e.,  $\zeta_{k_l} = \zeta$. Since UPA is a better choice than ULA for the proposed HB as analyzed in Section \ref{subsec:rf_chains} and verified in Section \ref{subsec:results_hb}, only UPA is considered in this subsection
. Specifically, the $16\times 16$ UPA with $\phi\in[\pi/6,5\pi/6]$ and $\theta\in[5\pi/9,3\pi/4]$ is considered in this subsection with its $R$ value that can provide the performance close to complete digital beamforming with $M$ RF chains, i.e., $R=64$.
In Fig. \ref{fig:results_upa_sc35}, the downlink average achievable sum rates of the proposed HB based on the proposed SC method with different $\gamma$ and $\zeta$ values under different SNRs in a TDD massive MIMO-OFDM system are presented. In the figure, the result of the proposed HB based on perfect SCM is presented as the reference curve, denoted by Perfect SCM. As for the proposed SC method, it is denoted by SC in the figure.

Based on Fig. \ref{fig:results_upa_sc35}, several observations are listed below.
\begin{enumerate}
\item{When $\gamma=\infty$ and $\zeta=1$, SC achieves almost the same performance as Perfect SCM, which indicates that the proposed SC algorithm can excellently track the spatial information of the second-order SCM and it is perfectly suited for the proposed HB method.}
\item{With perfect partial instantaneous CSI, i.e., $\zeta=1$, the result of SC with $\gamma=10$ is almost the same as the result of $\gamma=\infty$, which indicates that the dimension reduction method discussed in Section \ref{subsec:dimension} is valid.} 
\item{Even with $\zeta=0.1$, which implies that the CSI used to estimate the SCM is very inaccurate, SC can still achieve more than $97\%$ performance of Perfect SCM, which verifies the robustness of the proposed SC algorithm as analyzed in Section \ref{subsec:robustness}. Note that similarly to Section \ref{subsec:results_hb}, in all cases, perfect effective CSI seen at the baseband for digital beamforming is assumed to focus on the robustness of the proposed SC method.} 
\end{enumerate} 

\subsection{Application in FDD Systems} \label{subsec:results_fdd}
\ifCLASSOPTIONonecolumn
\begin{figure}[!t]
\centering \includegraphics[width = 1.0\linewidth]{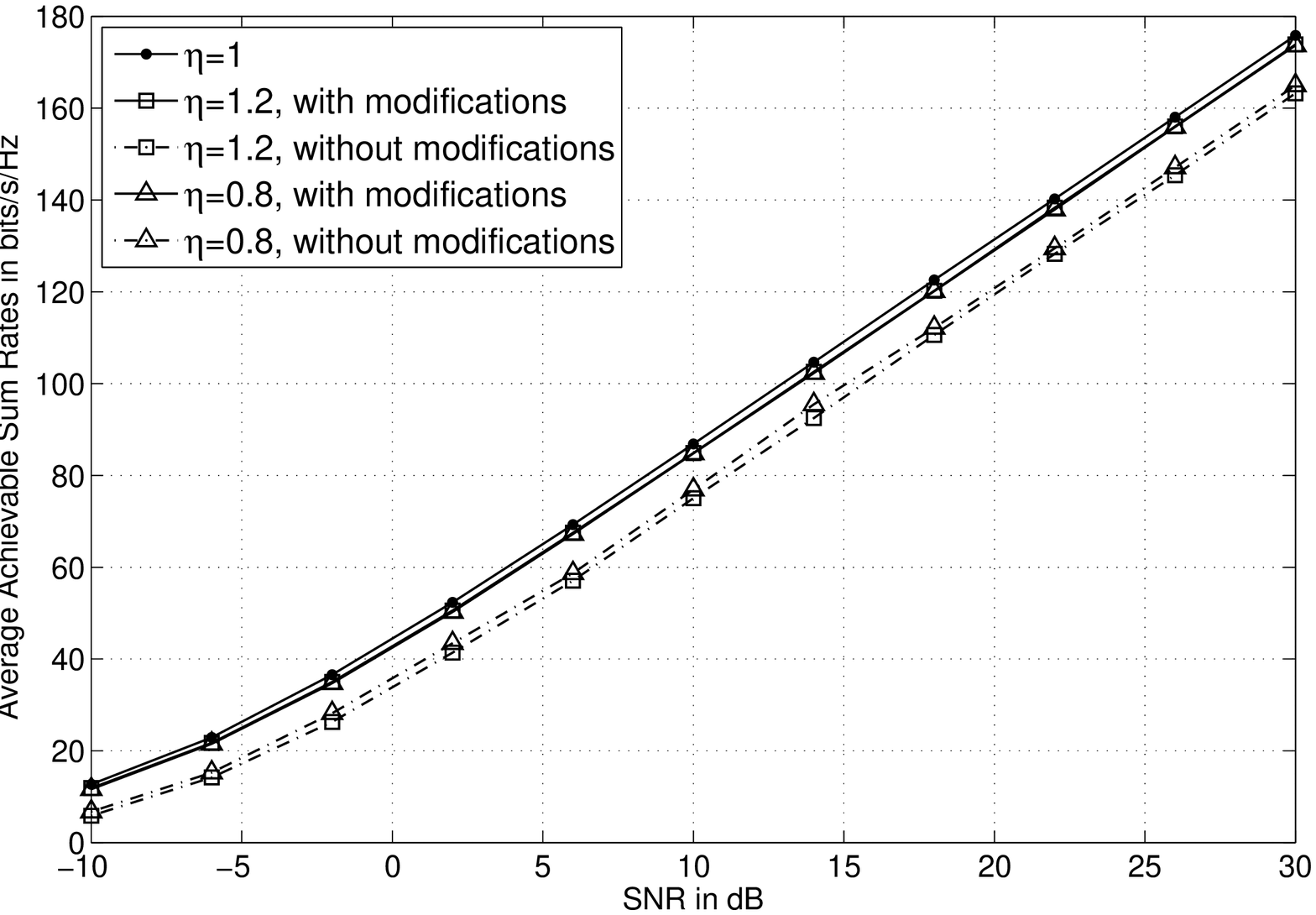}
\caption{Downlink average achievable sum rates of the proposed HB method based on the proposed SC algorithm for a $16\times 16$ UPA with $\phi\in[\pi/6,5\pi/6]$, $\theta\in[5\pi/9,3\pi/4]$, $\gamma=10$, $\zeta=1$, and $R=64$, employing different $\eta$ values under different SNRs in a FDD system.} 
\label{fig:results_upa_sc35_fdd}
\end{figure}
\else
\begin{figure}[!t]
\centering \includegraphics[width = 0.95\linewidth]{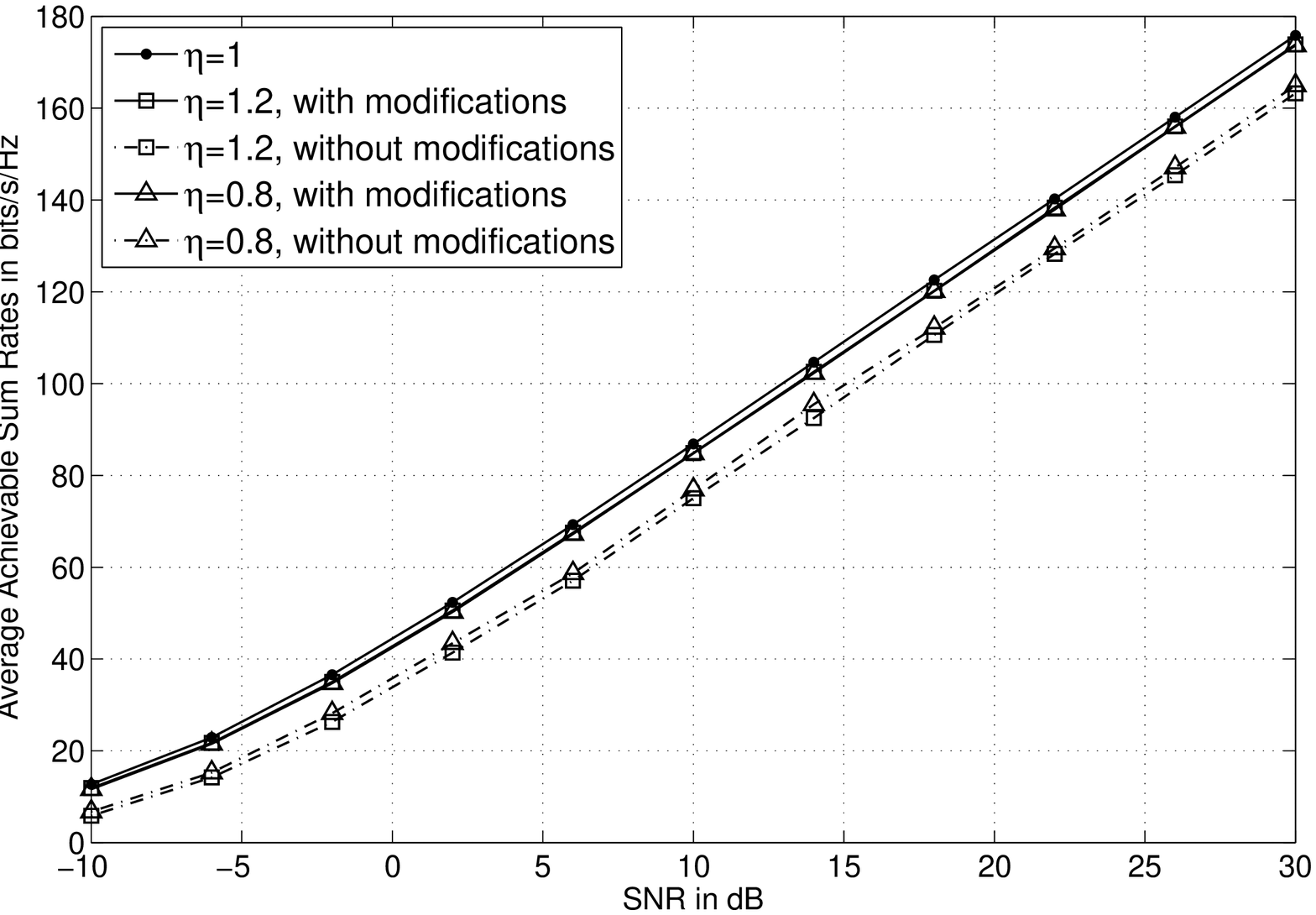}
\caption{Downlink average achievable sum rates of the proposed HB method based on the proposed SC algorithm for a $16\times 16$ UPA with $\phi\in[\pi/6,5\pi/6]$, $\theta\in[5\pi/9,3\pi/4]$, $\gamma=10$, $\zeta=1$, and $R=64$, employing different $\eta$ values under different SNRs in a FDD system.} 
\label{fig:results_upa_sc35_fdd}
\end{figure}
\fi


In this subsection, the performance of the proposed SC method in FDD systems presented in Section \ref{sec:fdd} is investigated. Note that since the same downlink carrier frequencies for TDD and FDD systems are considered in simulations, the application methods of the proposed SC algorithm presented in Section \ref{subsec:fdd_dcsi} based on partial downlink CSI achieve the same performance as TDD systems investigated in Section \ref{subsec:results_sc} in the case of perfect feedback. In practice, imperfect feedback results in performance loss. Since the feedback scheme that can provide the performance close to perfect feedback is a separate topic from the investigated SC algorithm, imperfect feedback is not considered in simulations. As a result, only the application method proposed in Section \ref{subsec:fdd_ucsi} based on partial uplink CSI is considered for simulations in this subsection. Based on the results presented in Section \ref{subsec:results_sc}, in simulations, the threshold $\gamma$ defined in (\ref{eq:threshold}) is selected as $\gamma = 10$, and perfect partial CSI used to estimate the SCM, i.e., $\zeta_{k_l} = \zeta = 1$ in (\ref{eq:est_csi_horizontal}), is assumed for all UEs. Similarly to Section \ref{subsec:results_sc}, every $1$ out of $48$ subcarriers, hence $25$ subcarriers in total,
are used for the expectation in (\ref{eq:cov_horizontal}), and only UPA is considered in this subsection. The UPA case simulated in Section \ref{subsec:results_sc} is also considered in this subsection. 
 In Fig. \ref{fig:results_upa_sc35_fdd}, the downlink average achievable sum rates of the proposed HB based on the proposed SC method for the considered UPA case is presented with different $\eta$ values where $\eta = F_{\mathrm{c}}^\mathrm{UL}/F_{\mathrm{c}}^\mathrm{DL}$ 
under different SNRs in a FDD massive MIMO-OFDM system. In the figure, the results with and without modified downlink steering angles based on (\ref{eq:dl_ul_horizontal}) and  (\ref{eq:dl_ul_vertical}) are both shown. Note that the result for $\eta = 1$ is just the reference curve that achieves the same performance as a TDD system investigated in Section \ref{subsec:results_sc}, which is in fact not achievable in a practical FDD system.

Based on Fig. \ref{fig:results_upa_sc35_fdd}, several observations are listed below.
\begin{enumerate}
\item{Without modifications, both $\eta=1.2$ and $\eta = 0.8$ suffer performance losses compared to $\eta =1$.}
\item{With modifications, both $\eta = 1.2$ and $\eta = 0.8$ can achieve the performance very close to $\eta = 1$, which verifies the effectiveness of the application method of the proposed SC algorithm presented in Section \ref{subsec:fdd_ucsi} based on partial uplink CSI.}
\item{Without modifications, although performance losses of $\eta=1.2$ and $\eta = 0.8$ compared to $\eta = 1$ are caused, decent performance can still be achieved, indicating that the proposed SC method without modifications could be directly applied to practical FDD systems, when the difference between the uplink and downlink carrier frequencies is not too large, with acceptable performance.} 
\end{enumerate}

\subsection{Discussion} \label{subsec:discussion}
\ifCLASSOPTIONonecolumn
\begin{figure}[!t]
\centering \includegraphics[width = 1.0\linewidth]{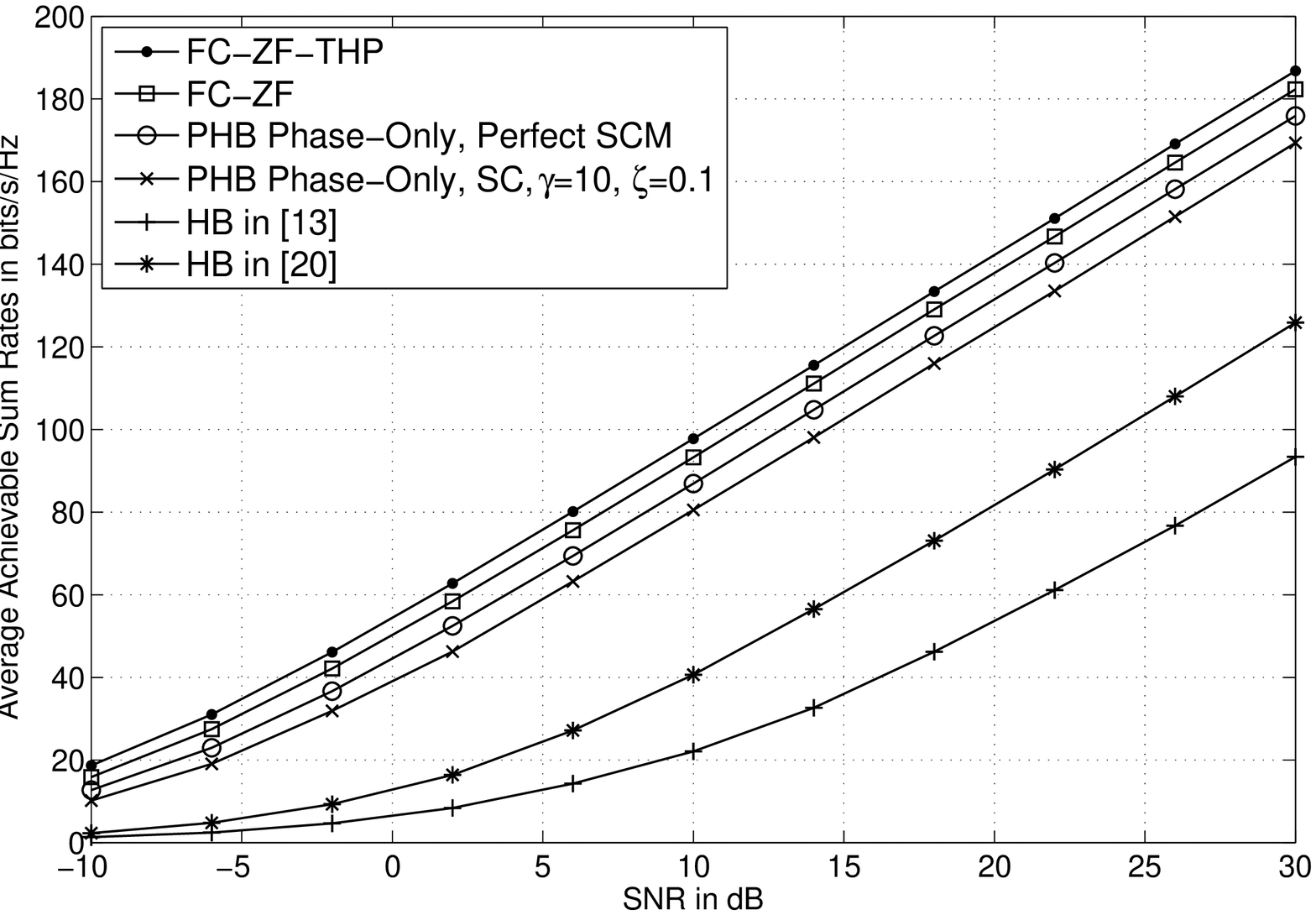}
\caption{Downlink average achievable sum rates of different HB methods with $R=64$ for a $16\times 16$ UPA with $\phi\in[\pi/6,5\pi/6]$ and $\theta\in[5\pi/9,3\pi/4]$, under different SNRs in a TDD system.} 
\label{fig:results_upa_com35}
\end{figure}
\else
\begin{figure}[!t]
\centering \includegraphics[width = 0.95\linewidth]{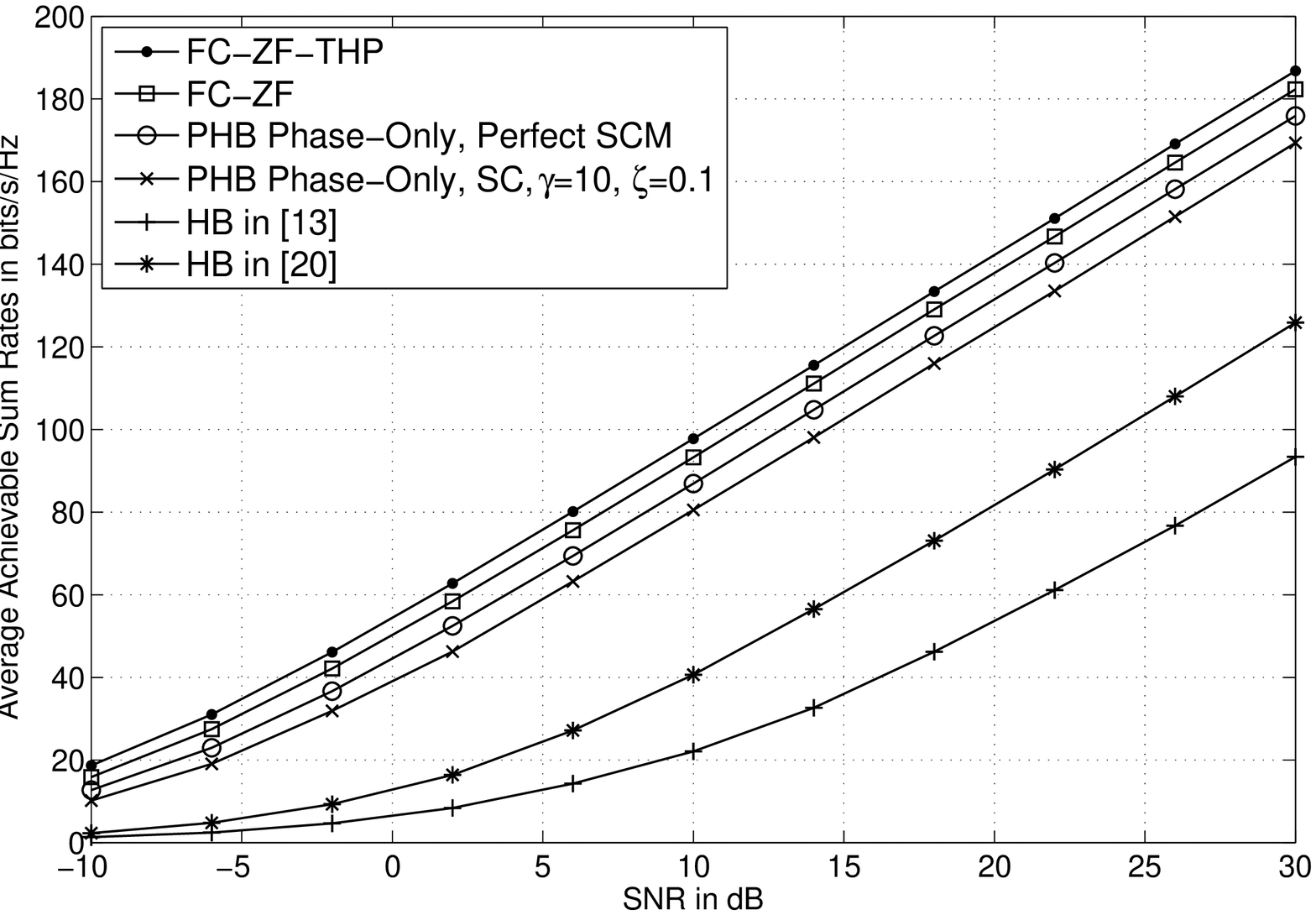}
\caption{Downlink average achievable sum rates of different HB methods with $R=64$ for a $16\times 16$ UPA with $\phi\in[\pi/6,5\pi/6]$ and $\theta\in[5\pi/9,3\pi/4]$, under different SNRs in a TDD system.} 
\label{fig:results_upa_com35}
\end{figure}
\fi

As explained in Section \ref{sec:introduction}, previous works on HB \cite{Wu_HB,Ayach_mmWave,Alkhateeb_HB_CE,Geng_MUHB,Liang_HB,
Bogale_MUHB,Xu_HB,Chiu_HB,Ying_HB,Stirling_MUHB,Alkhateeb_HP,Sohrabi_HB,Sohrabi_HB_OFDM} are not well suited for the considered application in this paper, i.e., a massive MU-MIMO OFDM system in sub $6\mathrm{GHz}$ frequency bands that serves multiple groups of UEs with each group being served by a different frequency resource. 
Specifically, HB methods in \cite{Ayach_mmWave,Alkhateeb_HB_CE,Wu_HB,Chiu_HB,Stirling_MUHB,Alkhateeb_HP} 
were designed for mmWave systems based on the extremely high dominance of the LOS component and the very limited number of multipath components in the channel model, which is not applicable to the channel model considered in the sub $6\mathrm{GHz}$ case 
where NLOS components are not negligible and the number of multipath components are generally not small. The HB methods presented in \cite{Geng_MUHB,Liang_HB,Xu_HB,Bogale_MUHB,Ying_HB,Sohrabi_HB,Sohrabi_HB_OFDM} can be employed in sub $6\mathrm{GHz}$ frequency bands. Unfortunately, they were specifically designed for a single UE or a single group of UEs all served on the same frequency resource. In the considered case 
that multiple groups of UEs are severed by different frequency resources, they cannot work properly. The reason is that the first-level AB is frequency-flat, so only one group of UEs can be properly served. To illustrate it, Fig. \ref{fig:results_upa_com35} compares the downlink average rates of the proposed HB method designed for multiple group of UEs and HB methods in \cite{Liang_HB} and \cite{Sohrabi_HB} designed for a single group of UEs, with $R=64$ for the same UPA case considered in Section \ref{subsec:results_sc} and Section \ref{subsec:results_fdd} under different SNRs in a TDD massive MIMO-OFDM system. Note that as shown in \cite{Liang_HB} and \cite{Sohrabi_HB}, their HB methods can achieve sum rates close to FC-ZF for a single group of UEs. However, as expected, Fig. \ref{fig:results_upa_com35} shows that when multiple groups of UEs are severed, they 
do not work properly, and the performance gaps compared to the proposed HB method 
are significant. Note that the HB methods in \cite{Liang_HB} and \cite{Sohrabi_HB} both require perfect full CSI, which is not required for the proposed HB method.  

\section{Conclusions} \label{sec:conclusions}
In this paper, focused on massive MIMO-OFDM systems in sub $6\mathrm{GHz}$ 
frequency bands, a novel HB algorithm with unified AB based on the second-order 
SCM is proposed to support multiple groups of UEs. The sufficient number of RF chains is derived for the proposed HB to achieve the performance close to complete digital beamforming without a reduced number of RF chains. Simulation results verify that with the proposed HB
method, the system cost of a massive MIMO-OFDM system can be greatly reduced with the performance loss 
no more than $5\%$. In addition, a novel practical SC algorithm is proposed 
to estimate the SCM required by the proposed HB. Simulation results show that the proposed SC method only suffers performance loss no more than $3\%$ compared to the perfect SCM case, and verify its high robustness against inaccurate CSI. The proposed methods can be applied to many scenarios. Particularly, its application in both TDD and FDD systems are discussed in details. Other than that, it is also applicable to both OFDM and single-carrier systems, both multiple groups of UEs and a single UE or a single group of UEs, as well as both LoS and NLoS channels. The generalization of proposed methods to the case that the AB is realized by a partially connected PSN, and to mmWave systems are considered in future work.  


%


\appendix [Derivations of (\ref{eq:dominant_search_fft})]
Because $\mathbf{R}^\mathrm{h}_{k_l}$ is a Hermitian matrix, based on (\ref{eq:vector_steering}), the objection function in (\ref{eq:dominant_search}) can be expanded as
\ifCLASSOPTIONonecolumn
\begin{align}
\omega^\mathrm{h}_{0k_l} = \mathop {\arg\max} \limits_{\omega \in \left[0,2\pi \right)} \frac{1}{M_{\mathrm{h}}} \left[ \sum_{m=1}^{M_\mathrm{h}}r_{mmk_l} + \sum_{p=1}^{M_\mathrm{h}-1} \sum_{m=1}^{M_\mathrm{h}-p}r_{m\left(m+p\right)k_l}e^{-jp\omega} + \sum_{p=1}^{M_\mathrm{h}-1} \sum_{m=1}^{M_\mathrm{h}-p}r_{\left(m+p\right)mk_l}e^{jp\omega} \right].
\label{eq:dominant_search_expand}
\end{align} 
\else
\begin{align}
\omega^\mathrm{h}_{0k_l} = & \mathop {\arg\max} \limits_{\omega \in \left[0,2\pi \right)} \frac{1}{M_{\mathrm{h}}} \left[ \sum_{m=1}^{M_\mathrm{h}}r_{mmk_l} \right. \nonumber \\
& + \sum_{p=1}^{M_\mathrm{h}-1} \sum_{m=1}^{M_\mathrm{h}-p}r_{m\left(m+p\right)k_l}e^{-jp\omega} \nonumber \\
& \left. + \sum_{p=1}^{M_\mathrm{h}-1} \sum_{m=1}^{M_\mathrm{h}-p}r_{\left(m+p\right)mk_l}e^{jp\omega} \right].
\label{eq:dominant_search_expand}
\end{align} 
\fi
Define a function $g(\omega)$ as 
\begin{align}
g\left(\omega\right) = \frac{1}{2} \sum_{m=1}^{M_\mathrm{h}}r_{mmk_l} + \sum_{p=1}^{M_\mathrm{h}-1} \sum_{m=1}^{M_\mathrm{h}-p}r_{m\left(m+p\right)k_l}e^{-jp\omega},
\label{eq:g_function}
\end{align} 
then (\ref{eq:dominant_search_expand}) can be rewritten as 
\begin{align}
\omega^\mathrm{h}_{0k_l} 
 = \mathop {\arg\max} \limits_{\omega \in \left[0,2\pi \right)} \frac{1}{M_{\mathrm{h}}} \left[ g\left(\omega\right) + g^*\left(\omega\right)\right] = \mathop {\arg\max} \limits_{\omega \in \left[0,2\pi \right)} \left\lbrace \Re\left[g\left(\omega\right) \right] \right\rbrace.
\label{eq:dominant_search_expand_g}
\end{align} 
Let the search step of $\omega$ as $2\pi/N_{\mathrm{F}}^{\mathrm{h}}$, then for the $n$th search, $n=1,\ldots,N_{\mathrm{F}}^{\mathrm{h}}$, 
\begin{align}
g\left(\omega = \frac{2\pi \left[n-1\right]}{N_{\mathrm{F}}^{\mathrm{h}}}\right) = \left(\mathbf{f}_{N_{\mathrm{F}}^{\mathrm{h}}}^n\right)^{\mathrm{T}} \boldsymbol{\rho}_{k_l}^\mathrm{h} = \mathcal{F}\left\lbrace \boldsymbol{\rho}_{k_l}^\mathrm{h} \right\rbrace^{N_\mathrm{F}^{\mathrm{h}}}_n,
\label{eq:g_function_fft}
\end{align}
where $\boldsymbol{\rho}_{k_l}^\mathrm{h}$ is defined in (\ref{eq:vector_spectial}), and $\mathbf{f}_{N_{\mathrm{F}}^{\mathrm{h}}}^n$ is defined as 
\begin{align}
\mathbf{f}_{N_{\mathrm{F}}^{\mathrm{h}}}^n = \left[1 \,\, e^{-j2\pi n/N_\mathrm{F}^{\mathrm{h}}} \,\, \cdots \,\, e^{-j2\pi n \left(N_\mathrm{F}^{\mathrm{h}} - 1 \right)/N_\mathrm{F}^{\mathrm{h}}} \right]^\mathrm{T}.
\label{eq:vector_fft}
\end{align}
As a result, based on (\ref{eq:dominant_search_expand_g}) and (\ref{eq:g_function_fft}), the final search problem (\ref{eq:dominant_search_fft}) is derived.
 
\section*{Acknowledgment}
The authors would like to thank the editor and the anonymous reviewers whose valuable comments improved the quality of the paper.


\ifCLASSOPTIONcaptionsoff
  \newpage
\fi

\bibliographystyle{IEEEtran}
\bibliography{IEEEabrv,Mybib}

\end{document}